\documentclass{elsart}

\usepackage{graphicx}
\usepackage{epsfig}

\usepackage{amssymb}

\begin{document}

\begin{frontmatter}



\title{Four Lectures on the \\ Physics of Crystal Growth}


\author{Joachim Krug}

\address{Fachbereich Physik, Universit\"at Essen, 45117 Essen, Germany}

\begin{abstract}
Several aspects of the theory of epitaxial 
crystal growth from atomic or molecular beams are developed from 
the perspective of statistical physics. Lectures are devoted
to the rate equation theory of two-dimensional nucleation and its limitations; 
the growth of multilayer wedding cakes
in the presence of strong step edge barriers; the continuum theory of mound
coarsening; and growth-induced step meandering on vicinal surfaces.  
\end{abstract}

\end{frontmatter}

\section{Introduction}

The growth of a crystalline film from a molecular or atomic beam, 
commonly referred to as Molecular Beam Epitaxy (MBE), is a simple
example of a self-assembly process. In contrast to crystallization from the
melt, which often leads to dendrites and other ramified patterns
\cite{Cummins89}, MBE growth can be described without reference to the
transport of matter, latent heat or impurities in the fluid phase. 
The remarkable richness of patterns forming during MBE is determined
solely by processes which occur locally at the surface. Moreover,
in the case of \emph{homoepitaxial} growth, in which a film is grown on a 
substrate of the same material,  energetic determinants such as 
interfacial free energies and misfit strain are absent, so that the
film morphology is governed primarily by growth kinetics. This makes
homoepitaxy an ideal laboratory in which to study the emergence of
mesoscopic patterns and their associated length scales through
self-organization processes far from equilibrium. An additional benefit
is that, since the invention of scanning tunneling microscopy, 
this laboratory is open to direct visual inspection. 

The main steps in the growth process can be summarized as 
follows\footnote{Throughout these notes, the unit of length will be the
substrate lattice spacing. Thus the deposition rate $F$ denotes the number
of atoms deposited per unit time and adsorption site, and the diffusion
coefficients $D$ and $D'$ are actually hopping rates. All three quantities
have units of inverse time.}.
Atoms are deposited at rate $F$. They migrate along the surface
with a two-dimensional diffusion coefficient $D$. When two atoms meet
they form a dimer. Dimers may subsequently disintegrate, or they may
grow by aggregation of further atoms into trimers and larger clusters.
Once a substantial fraction of the surface is covered by two-dimensional
island clusters, these begin to coalesce and a full atomic layer forms,
on which the processes involved in producing the first layer repeat
themselves. 

\begin{figure}[htb]
\centerline{\epsfig{figure=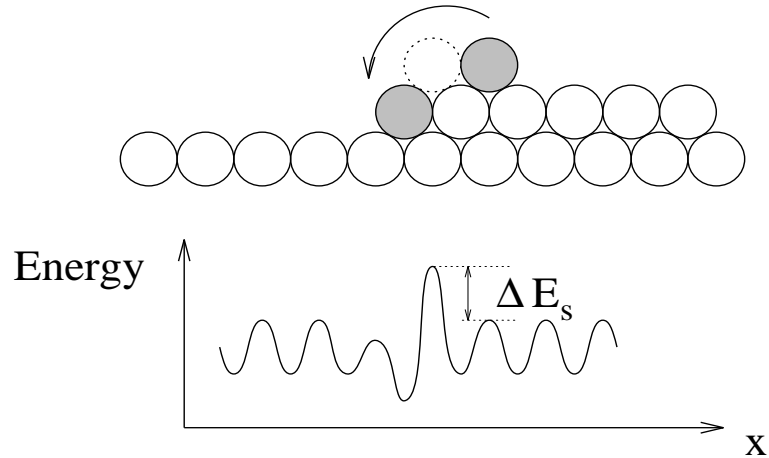,height=6cm,width=10cm,angle=0}}
\caption[]{An atom descending from a step edge experiences an additional
energy barrier $\Delta E_{\mathrm{S}}$.}
\label{esfig}
\end{figure}

At this stage of the growth process, the fate of atoms deposited
on top of first layer islands becomes important. Such atoms may either
descend from the island, thus contributing to the growth of the island
edge, or they may remain on the island, promoting the nucleation of the
next layer. On many crystal surfaces, the diffusion of atoms between different
atomic layers is suppressed due to additional energy barriers,
which an atom crossing a step has to overcome. This phenomenon
was first observed experimentally by Ehrlich and Hudda \cite{Ehrlich66},
and some of its consequences for the growth of stepped surfaces were
analyzed by Schwoebel and Shipsey \cite{Schwoebel66}. The atomistic
origin of the additional step edge barrier is illustrated in Figure 
\ref{esfig}: 
An atom descending from a step edge passes through
a transition state of very low coordination, which implies poor binding
and thus a higher energy. This picture is oversimplified, because in many
cases descent by concerted
exchange is more facile than hopping \cite{Feibelman98}. 
Nevertheless, it 
is generally true that the rate for interlayer diffusion, in the following
denoted by $D'$, is smaller than the in-layer diffusion constant $D$.

In-layer and interlayer diffusion, as well as all other atomic processes
involved in the growth, decay and shape changes of two-dimensional
islands, are \emph{thermally activated}. Let us recall what this means,
using the in-layer migration process as illustration. To a good approximation,
the motion of an adsorbed atom (an \emph{adatom}) on a crystal surface
can be viewed as a two-dimensional random walk between adsorption sites.
In hopping from one adsorption site to another, the adatom has to overcome
an energy barrier $E_D$. The energy is provided by thermal substrate
vibrations. This implies the familiar Arrhenius form
\begin{equation}
\label{Arr}
D = D_0 \, e^{-E_D/k_B T}
\end{equation}
for the diffusion coefficient. In (\ref{Arr}), $T$ is the substrate
temperature, $k_B$ the Boltzmann constant, and $D_0$ is an attempt
frequency with a typical magnitude around $10^{13} \mathrm{s}^{-1}$. 
The main role of temperature in MBE growth is that it regulates, through
expressions like (\ref{Arr}), the relative rates of the different
activated processes on the surface. 

\begin{figure}
\centerline{\includegraphics[width=0.95\textwidth]{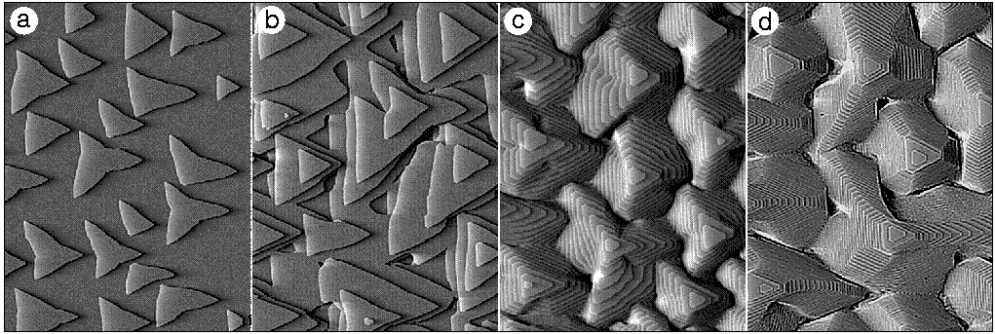}}
\caption{Growth of Pt on Pt(111) at $T = 440 \mathrm{K}$  
\protect\cite{Kalff99b}. 
The total
coverage is (a) 0.3 monolayers (ML), (b) 3 ML, (c) 12 ML and
(d) 90 ML. The image size is $2600 {\mathrm{\AA}} \times 3450 {\mathrm{\AA}}$.
Courtesy of T. Michely.}
\label{PtMounds}
\end{figure}

In these lectures we will explore some of the pattern forming phenomena
that arise through the interplay of the three kinetic rates 
$D$, $D'$ and $F$. In Section \ref{Nucleation}, the classical atomistic
rate equation theory of two-dimensional nucleation will be briefly reviewed.
Its central result is a scaling law, Eq.(\ref{Nscal}), which relates the
spacing between first layer island clusters to the ratio $D/F$. The 
limitations of the classical theory become evident in the treatment of
nucleation on top of islands, which requires statistical arguments beyond
rate equations. Sections \ref{Wedding} and \ref{Continuum} are devoted
to the mound patterns which appear generically in multilayer growth
(see Figure \ref{PtMounds} for an example).
The approach taken in Section \ref{Wedding} is quite atomistic, focusing
on the distribution of the deposited mass among the different layers and
the corresponding mound shapes. Section \ref{Continuum} provides
a more macroscopic perspective, which allows one to address also the 
mass transport between mounds and the coarsening of the pattern. 
The final Section \ref{Vicinals} is devoted to growth on vicinal surfaces,
which are intrinsically anisotropic due to the presence of preexisting steps.

The treatment of these topics is far from exhaustive. More details can
be found in several recent books and review articles 
\cite{Politi00,Pimpinelli98,Markov95,Saito96,Venables00,Krug97a,Barabasi95}. 
The lectures do not cover the theory of kinetic roughening by stochastic
fluctuations, which in fact initiated, through the seminal work of 
Kardar, Parisi and Zhang \cite{Kardar86}, the interest of the statistical
physics community in film growth. The reasons for this omission are twofold.
First, the subject has been extensively reviewed elsewhere 
\cite{Krug97a,Barabasi95,Krug91,Halpin95}. Second, despite its tremendous 
conceptual impact, the relevance of kinetic roughening theory to real film 
growth has still not been quantitatively demonstrated\footnote{Experimental 
work until 1995 has been reviewed in \cite{Palasantzas95}.}. This is in 
contrast to the growth instabilities discussed in the present lectures, where
a quantitative linking of mesoscopic patterns to specific atomistic processes 
appears to be well within reach. A possible manifestation of kinetic
roughening in a setting which is relevant to MBE growth is the noise-induced 
damping of growth oscillations in layer-by-layer growth. Here kinetic roughening theory has been used to establish scaling relationships between the damping time and 
the ratio $D/F$ \cite{Krug97a,Kallabis97,Rost97a,Ross00}.

\section{Two-dimensional nucleation}
\label{Nucleation}

The formation of an atomic layer on a high symmetry substrate
without steps or defects has to proceed through the congregation
of mobile adatoms into stable clusters, which subsequently
grow by accretion. The process is analogous to the nucleation of
a condensed phase out of a
supersaturated gas, as described
by thermodynamic nucleation theory
\cite{Volmer39}. The central object of this
theory is the {\em critical nucleus}, which defines the free energy barrier
that has to be surmounted to reach the stable phase. The size of the
critical nucleus is inversely proportional to the supersaturation.

In far from equilibrium growth the critical nucleus size may
reach atomic dimensions, thus precluding a straightforward
application of thermodynamic concepts and necessitating the development
of an {\em atomistic} theory of nucleation kinetics \cite{Walton62}.
This was achieved in the 1960's and 70's by Zinsmeister, Stowell,
Venables and others;
extensive reviews are available \cite{Venables73,Stoyanov81,Venables84}.
Precise experimental tests of the theory have become
possible only recently \cite{Brune98,Brune99}. In addition,
large scale computer simulations play an
increasingly important r\^ole in establishing the
validity and limitations of nucleation theory, as they
allow for separate scrutiny of the various assumptions
going into the theory.

The atomistic theory of two-dimensional nucleation is summarized in 
the next section. We then turn to the problem of second layer nucleation on top
of islands, which differs from the nucleation of the first layer because
of the confinement of the atoms by step edge barriers. This problem is of 
conceptual interest because it illustrates the limitations of mean field
rate equation theory \cite{Krug00a,Heinrichs00,Krug00b}. 
At the same time a quantitative theory of second
layer nucleation is a necessary prerequisite for the discussion of multilayer
growth in the subsequent lectures.

\subsection{Rate equation theory}

\label{Rateeq}

The classical approach to nucleation kinetics starts from
balance or rate equations for the areal concentrations $n_s$ of clusters
consisting of $s$ atoms; $n_1$ is the adatom density,
$n_2$ the density of dimers, and so on.
To be precise, we define $n_s$ as the
number of clusters per surface area, averaged over a region
containing a large number of clusters. If the adatoms are
the only mobile species (the mobility of larger
clusters is negligibly small), then clusters grow
solely by aggregation of single adatoms. Defining $\Gamma_s$ to
be the net rate at which $s+1$-clusters form from $s$-clusters,
we have for $s \geq 2$
\begin{equation}
\label{ratenj}
\frac{d n_s}{dt} = \Gamma_{s-1} - \Gamma_{s} \;\;\;\;\;(s€\geq 2).
\end{equation}
The net formation rates $\Gamma_s$ can be written as
\begin{equation}
\label{Gammaj}
\Gamma_s = \sigma_s D n_1 n_s - \gamma_{s+1} n_{s+1},
\end{equation}
where $\gamma_s$ is the rate at which adatoms detach from
a $s$-cluster and the dimensionless {\em capture number} $\sigma_s$
accounts for the propensity of a $s$-cluster to absorb an adatom (see below).
The chain (\ref{ratenj}) of aggregation equations is fed by the
adatom density $n_1$. If desorption from the surface can be neglected
(the {\em complete condensation} limit \cite{Venables73,Venables84}),
adatoms are lost only through dimer formation and capture at
larger clusters, and the adatom rate equation reads
\begin{equation}
\label{n1}
\frac{dn_1}{dt} = F - 2 \Gamma_1 - \sum_{s \geq 2} \Gamma_s.
\end{equation}
The deposition rate $F$ is defined as the number of atoms arriving
per unit time and surface area.

In principle, Eqs.(\ref{ratenj}-\ref{n1}) provide a complete description
of the nucleation process; in practice, they contain far too many
(generally unknown) kinetic parameters to be useful.
This difficulty is commonly circumvented
by introducing a distinction between stable and unstable
clusters, and postulating a separation of time scales between the
kinetics of the two kinds. Stable clusters of sizes
$s \geq i^\ast + 1$ are assumed not to decay,
i.e. $\gamma_s = 0$ for $s \geq i^\ast$,
while the detachment of adatoms from
unstable clusters with $s \leq i^\ast$ occurs
sufficiently rapidly to establish thermodynamic equilibrium
between the different size populations. It is important to note that,
in contrast to thermodynamic nucleation theory, the critical cluster size
$i^\ast$ introduced here
contains a {\em kinetic} element, as it refers to stability
and equilibration only on the time scale
relevant to the deposition experiment.

Next the total density $N$ of stable clusters, also referred to as
{\em islands} in the following, is introduced through
\begin{equation}
\label{Ndef}
N = \sum_{s = i^\ast+1}^\infty n_s.
\end{equation}
Summing (\ref{ratenj}) from $s=i^\ast + 1$ this is seen to evolve
according to
\begin{equation}
\label{Ntime}
\frac{dN}{dt} = \sigma_{i^\ast} D n_1 n_ {i^\ast}.
\end{equation}
The assumption of thermal equilibrium among unstable clusters implies that
the net formation rates $\Gamma_s$ vanish for $1 \leq s \leq i^\ast - 1$, and
induces the detailed balance relations \cite{Walton62}
\begin{equation}
\label{Walton}
n_s \approx n_1^s \mathrm{e}^{E_s/k_{\mathrm B} T} 
\;\;\;\;\; (2 \leq s \leq i^\ast)
\end{equation}
between the unstable cluster concentrations and the adatom concentration.
Here $E_s$ is the total (positive)
binding energy of an $s$-cluster,
i.e. the energy needed to disperse the cluster
into single adatoms; note
that $E_1 = 0$.

Using (\ref{Walton}) the nucleation rate on the right hand side of (\ref{Ntime})
can be expressed in terms of the adatom density. To complete the description,
the rate equation for $n_1$ is simplified by introducing the
average capture number for stable islands
\begin{equation}
\label{sigmabar}
\bar{\sigma} = N^{-1} \sum_{s = i^\ast + 1}^\infty n_s \sigma_s.
\end{equation}
Equation (\ref{n1}) can then be written in the form
\begin{equation}
\label{n1b}
\frac{dn_1}{dt} = F - \sigma_{i^\ast} D n_1 n_{i^\ast} - \bar{\sigma} D n_1 N.
\end{equation}
Together (\ref{Ntime}), (\ref{Walton}) and (\ref{n1b}) form a closed set of equations
from which the island and adatom densities can be computed.

Before turning to the solution of these equations,
some remarks concerning the capture numbers $\sigma_s$ are in order.
In the early literature on atomistic nucleation theory, a geometric
view of the capture numbers as effective cross sections for
atoms colliding with clusters prevailed.
More appropriately, they are
defined through the requirement that the flux of adatoms to
a cluster of size $s$ should be, on average, equal to $\sigma_s D n_1$.
The capture number of a cluster evidently depends on its size, but
in general it depends on the sizes and locations of surrounding clusters
as well, because these affect the adatom concentration field. The calculation
of $\sigma_s$ involves the solution of a diffusion equation for the
adatom concentration, with appropriate boundary conditions representing both the
capture of adatoms at the cluster of interest, and the presence of other clusters
far away \cite{Bales94}. 
In this sense the capture numbers retain some information about the
spatial arrangement of the clusters, which is otherwise not accounted for
in the rate equations; in the jargon of statistical
mechanics, the rate equation description constitutes a
{\em mean field} approximation.
In practice, both
$\sigma_{i^\ast}$ and $\bar{\sigma}$
turn out to be slowly varying functions which can be replaced
by constants for many (though not all) purposes. We adopt this
simplification for the present discussion.

The solutions of the
coupled equations (\ref{Ntime}) and (\ref{n1b}) display
two temporal regimes. In the early time, {\em transient}
nucleation regime the loss terms on the right hand side
of (\ref{n1b}) are negligible. The adatom concentration
increases proportional to the total coverage $\Theta = F t$,
and the island density grows rapidly
as $N \sim \Theta^{i^\ast + 2}$. This regime ends when capture
of adatoms at stable islands becomes appreciable, at a coverage
which can be estimated by comparing the first and last terms
on the right hand side of (\ref{n1b}). In the subsequent
{\em steady state} regime these two terms balance completely,
and the adatom density is determined by the island
density through
\begin{equation}
\label{n1steady}
n_1 \approx \frac{F}{D \bar{\sigma} N}.
\end{equation}
Inserting this into (\ref{Ntime},\ref{Walton}) and
integrating in time yields the central result of nucleation
theory,
\begin{equation}
\label{Nsol}
N \approx \Theta^{1/(i^\ast + 2)}
\left(\frac{F}{D} \right)^{
\frac{i^\ast}{i^\ast + 2}} \mathrm{e}^{E_{i^\ast}/(i^\ast + 2)
k_{\mathrm{B}} T}.
\end{equation}
The most important feature of this expression
is that it takes the form of a \emph{scaling relation}
\begin{equation}
\label{Nscal}
N \sim \left(\frac{F}{D} \right)^{\chi}
\end{equation}
between the island number density and the
ratio of the two basic kinetic rates $D$ and $F$ of the
deposition process, with the \emph{scaling exponent} 
$\chi$ taking the value $\chi = i^\ast/(i^\ast + 2)$
(for a generalized expression see (\ref{chigen})).

The rate equations formulated above are restricted to the low
coverage regime, since finite coverage effects such as direct
impingement and cluster-cluster coalescence have been neglected.
In the absence of cluster mobility, coalescence provides the
only mechanism by which the island density can decrease.
This produces a maximum in $N(\Theta)$ which is often
used as a convenient reference point for
the experimental determination of cluster densities.

\subsection{The rate of second layer nucleation}

\label{2nd}

\begin{figure}
\centerline{\epsfig{figure=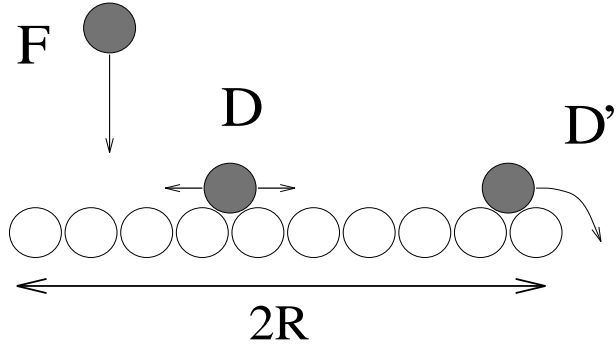,height=4.5cm,width=8.cm,angle=0}}
\caption{Kinetic processes involved in second layer nucleation. The figure
shows a circular island viewed from the side.}
\label{2ndfig}
\end{figure}

We now ask how the nucleation process is modified when it occurs
on top of an island, rather than on the (unbounded) substrate.
The geometry is illustrated in Figure \ref{2ndfig}. Atoms are
being deposited at rate $F$ onto a circular island of radius $R$. They
diffuse on the island with an in-layer diffusion constant $D$, and
descend from it with an (average) interlayer diffusion rate $D' < D$. 
We are looking for the probability per unit time, 
$\omega$, for a nucleation event to occur on the island. 
Assuming that dimers of adsorbed atoms are stable (as they will be
at sufficiently low temperatures), nucleation takes place as soon as
two adatoms meet. In the terminology of the preceding subsection,
this corresponds to $i^\ast = 1$; for a generalization of the following
considerations to $i^\ast > 1$ see \cite{Krug00b}. 

The kinetic rates $F$, $D$ and $D'$ combine to form three relevant
time scales: 
The \emph{interarrival time} 
\begin{equation}
\label{deltat}
\Delta t = \frac{1}{\pi R^2 F}
\end{equation}
between subsequent depositions onto the island; the \emph{diffusion time}
\begin{equation}
\label{tauD}
\tau_D \sim R^2/D
\end{equation}
required for an atom to diffuse once across the island, or for two atoms
to meet; and the \emph{residence time} $\tau$ which an atom spends on the
island before descending, when no nucleation event takes place. The calculation
of the residence time requires the solution of a stationary diffusion problem
with appropriate boundary conditions at the step edge, which yields 
\cite{Krug00a,Tersoff94}
\begin{equation}
\label{tau}
\tau = \frac{R^2}{8 D} + \frac{R}{2 D'}.
\end{equation}
The residence time is comparable to the diffusion time $\tau_D$ if the 
suppression of interlayer transport is weak, in the sense that
$D/D' \ll R$, while in the 
opposite regime of \emph{strong step edge barriers},
$D/D' \gg R$, the residence time becomes $\tau = R/2 D'$ independent
of $\nu$. 

We will focus on the latter regime in the following. Then 
$\tau \gg \tau_D$, which implies that two atoms are certain to meet once
they are present simultaneously on the island. The probability for this to occur
is $\tau/(\tau + \Delta t) \approx \tau/\Delta t$ if $\Delta t \gg \tau$,
which is true for reasonable deposition fluxes. Multiplying this by the
total number of atoms deposited onto the island per unit time, we obtain
the expression \cite{Krug00a}
\begin{equation}
\label{omega}
\omega = \frac{\tau}{(\Delta t)^2} = \frac{\pi^2 F^2 R^5}{2 D'}
\end{equation}
for the nucleation rate, which is \emph{exact} under the stated conditions.

\subsection{The limitations of mean field nucleation theory}

\label{NucRW}

The first calculation of 
the rate of second layer nucleation \cite{Tersoff94} was 
based on the rate equation (\ref{Ntime}),
in which the average 
adatom density $n_1$ is replaced by its local value 
$n(\vec{r})$ at a point $\vec{r}$ on the island. The
\emph{local} nucleation rate $I(\vec{r})$, which counts
the number of nucleation events per time and lattice
site, is then given by  
\begin{equation}
\label{I(r)}
I(\vec{r}) =  \sigma_{i^\ast} D e^{E_{i^\ast}/k_{\mathrm B} T}
n(\vec{r})^{i^\ast + 1}.
\end{equation}
Computing the adatom density profile from the stationary diffusion
equation and integrating over
the island area, the total nucleation rate on a circular island
is obtained. Comparison with the exact expression (\ref{omega}) 
for $i^\ast = 1$ shows that the
rate equation approach overestimates  
(\ref{omega}) by a factor of the order
of $D/RD' \gg 1$ \cite{Krug00a}. 
 
To elucidate the origin of this discrepancy, it
is useful to view 
the encounter of the two atoms as a first passage
problem in the four-dimensional space spanned by their coordinates. 
The joint diffusion of the atoms stops either
when the two reach adjacent lattice sites (nucleation) or
when one of them escapes, whatever happens first. The 
detailed analysis of this process \cite{Castellano01} 
reveals the nature of the approximation based
on (\ref{I(r)}): Setting the nucleation rate proportional to 
the square of the adatom density amounts to treating the atoms
as \emph{noninteracting}, in the sense that they are allowed to 
continue their diffusion even after they have met; in this
way a single pair can accumulate several (spurious) nucleation events,
and the nucleation rate is overestimated. 

For a quantitative comparison, 
we note that (for $i^\ast = 1$) the nucleation probability
$p_{\mathrm {nuc}}$ per atom can be written quite generally as the product of the
(mean) adatom density and the number 
$N_{\mathrm{dis}}(\tau)$ of \emph{distinct} sites the atom
encounters during its lifetime \cite{Villain92}. 
The validity of this statement becomes evident by
assuming that all other adatoms are immobile: Then it is clear
that repeated visits to the same unoccupied site do not increase
the chance for nucleation. The lifetime of the atom equals its
residence time in the case of second layer nucleation, but the
argument applies equally well to the nucleation of the first layer,
where the lifetime is determined by capture at stable islands, see below.

Using the general relation \cite{Krug00a}
\begin{equation}
\label{nbar}
\tau = \bar n/F
\end{equation}
between the residence time and the mean adatom density $\bar n$,
which follows from simple mass balance considerations, the
total nucleation rate on the island can then be written as 
\begin{equation}
\label{omega1}
\omega = (\Delta t)^{-1} p_{\mathrm {nuc}} = F N_{\mathrm{dis}} 
\frac{\tau}{\Delta t}.
\end{equation}
In the strong barrier limit, $N_{\mathrm{dis}}$ equals the total number
of sites on the island, and (\ref{omega1}) reproduces 
(\ref{omega}). To compare (\ref{omega1}) 
to the rate equation approximation,
we multiply the local nucleation rate (\ref{I(r)}) by the island
area and obtain, in order of magnitude,
\begin{equation}
\label{omegamf}
\omega_{\mathrm {mf}} \sim R^2 D (\bar n)^2 \sim F 
{N}_{\mathrm {all}} \frac{\tau}{\Delta t},
\end{equation}
where (\ref{nbar}) has been used, and the number 
${N}_{\mathrm {all}} = D \tau$ of \emph{all} sites
visited by an adatom during its residence has been introduced.
Thus the expressions (\ref{omega1}) and (\ref{omegamf}) differ,
in general, by a factor ${N}_{\mathrm {all}}/{N}_{\mathrm{dis}} > 1$. 

The distinction between $N_{\mathrm{all}}$ and $N_{\mathrm{dis}}$
affects also the mean field calculation of the density of first layer
islands in Section \ref{Rateeq} \cite{Villain92}. 
In this case the life time of
a freshly deposited adatom is of the order $\tau \sim 1/(N D)$, and
the average adatom density is $n_1 =  F \tau \sim F/N D$, compare to
(\ref{n1steady}). The theory of two-dimensional random walks
provides the expression ${N}_{\mathrm{dis}} \approx \pi D \tau/\ln(D \tau)$
for $D \tau \gg 1$ \cite{Itzykson89}. The actual value of $N$ is
fixed by the requirement that, out of the approximately $1/N$
adatoms deposited in the area occupied by one island, only one (or
two!) forms a nucleus, i.e. that $p_{\rm nuc} \sim
n_1 {N}_{\mathrm{dis}}(\tau) \sim N$. This yields finally
\begin{equation}
\label{Nlog}
{N}^3 \ln(1/N) \approx \frac{F}{D},
\end{equation}
which coincides with the scaling law
(\ref{Nscal}) for $i^\ast = 1$
only up to a logarithmic
correction; the leading behavior of (\ref{Nlog}) for
$D/F \gg 1$ is
\begin{equation}
\label{Nlog2}
N \sim (F/D)^{1/3}
[\log (D/F)]^{-1/3}.
\end{equation}
In a power law fit of $N$ versus
$F/D$ this will tend to produce
scaling exponents which are smaller than 1/3. 

The logarithmic factor can be reproduced within
rate equation theory
by using the expression
\begin{equation}
\label{sigma1}
\sigma_1 \approx \frac{4 \pi}{\ln[(D/F) n_1]}
\sim - \frac{1}{\ln( N)}
\end{equation}
for the capture number of adatoms \cite{Bales94}.
Reduced rate equations of the form (\ref{Ntime},\ref{n1b})
which incorporate the logarithmic correction
are able to quantitatively reproduce
the results of kinetic Monte Carlo simulations
\cite{Tang93}.

The random walk picture has also been helpful in clarifying
the case of one-dimensional
diffusion. In one dimension the distance
between islands is $N^{-1}$, and correspondingly the adatom
lifetime in the steady state regime is of the order of
$\tau \sim 1/(N^2 D)$ and the adatom density is $n_1 \sim
F/(N^2 D)$. The number of distinct sites visited by a one-dimensional
random walk grows as ${N}_{\mathrm{dis}}(\tau) \sim \sqrt{D \tau}$.
Inserting these expressions into the condition $p_{\rm nuc} \sim N$
yields the estimate $N \sim (F/D)^{1/4}$ for $i^\ast = 1$
\cite{Villain92}.
Similar considerations can be employed to derive
a useful general formula for the
exponent $\chi$ in the scaling law (\ref{Nscal})
\cite{Krug00b,Kallabis98}. It reads
\begin{equation}
\label{chigen}
\chi = \frac{d i^\ast}{d + 2 i^\ast + \min[d i^\ast,2]}.
\end{equation}
Here $d=1,2$ denotes the dimensionality of diffusion.
Logarithmic corrections similar to (\ref{Nlog}) arise whenever
$d i^\ast = 2$, because $d i^\ast$ is the effective
dimensionality of the random walk in configuration space
which describes the nucleation process, and
two-dimensional random walks are well known to be
marginally space filling \cite{Itzykson89}.

\section{Wedding cakes}
\label{Wedding}

It was first predicted by Villain \cite{Villain91} that the existence
of step edge barriers -- the fact that $D' < D$ -- implies a growth instability,
in which a mound morphology develops on a flat crystal surface. This phenomenon,
which was subsequently observed on a variety of substrates 
\cite{Ernst94,Johnson94,Thuermer95,Stroscio95,Tsui96,Zuo97},
is the subject of the following two lectures.

\subsection{Poisson growth}
\label{Poissonmodel}

To appreciate the relevance of interlayer transport for multilayer
growth, we first consider the case where it is completely absent,
i.e. we set $D' = 0$. Then
each adatom remains in the layer in which it was first deposited,
and is incorporated into that layer at an ascending step edge.
This implies a simple evolution of the layer coverages 
$\theta_n$, $0 \leq \theta_n \leq 1$, 
where $n = 1$, 2, 3...counts the layers and $\theta_0 = 1$ describes
the substrate. The rate at which layer $n$ grows is proportional
to the {\em exposed} coverage  
\begin{equation}
\label{cn}
\varphi_{n-1} = \theta_{n-1} - \theta_{n}
\end{equation}
of the layer $n-1$ below, and therefore the $\theta_n$ satisfy 
\cite{Cohen89}
\begin{equation}
\label{Poissongrowth}
\frac{d\theta_n}{dt} = F (\theta_{n-1} - \theta_n)
\end{equation}
with the initial conditions
$\theta_{n \geq 1}(t=0) = 0$. It is straightforward to check that
the solution reads 
\begin{equation}
\label{Poisson}
\theta_n = 1 - e^{-\Theta} \sum_{k=0}^{n-1} \frac{\Theta^k}{k !},
\end{equation}
where
\begin{equation}
\label{Total}
\Theta = \sum_{n=1}^\infty \theta_n = Ft
\end{equation}
is the total deposited coverage. Correspondingly the 
exposed coverages follow a Poisson distribution with parameter
$\Theta$, 
\begin{equation}
\label{Poisson2}
\varphi_n = \frac{e^{-\Theta} \Theta^n}{n !}.
\end{equation}

Since $\varphi_n$ is the probability that an arbitrary point on the surface
resides on layer $n$, it can also be viewed as the probability
distribution of the local film height, measured in units of the layer
thickness. The mean height is $\Theta$, and the standard deviation of 
the $\varphi_n$ defines the {\em surface width} $W$, a common measure
of film roughness, through
\begin{equation}
\label{W}
W^2 = \sum_{n=0}^\infty (n - \Theta)^2 \varphi_n.
\end{equation}
For the Poisson distribution (\ref{Poisson2}), the variance is
equal to the mean, and therefore 
\begin{equation}
\label{Wstat}
W = \sqrt{\Theta}
\end{equation}
for growth without interlayer transport. 
Remarkably, the expression (\ref{Wstat}) is independent of the
in-layer diffusion rate. It represents the maximum surface
roughness that can be generated by the randomness in the
deposition flux, and is referred to in the literature
as the {\em statistical growth} or {\em random deposition}
limit \cite{Barabasi95}. Interlayer transport is solely
responsible for reducing the roughness below this limit. 

\begin{figure}[htb]
\begin{center}
\includegraphics[width=0.48\textwidth]{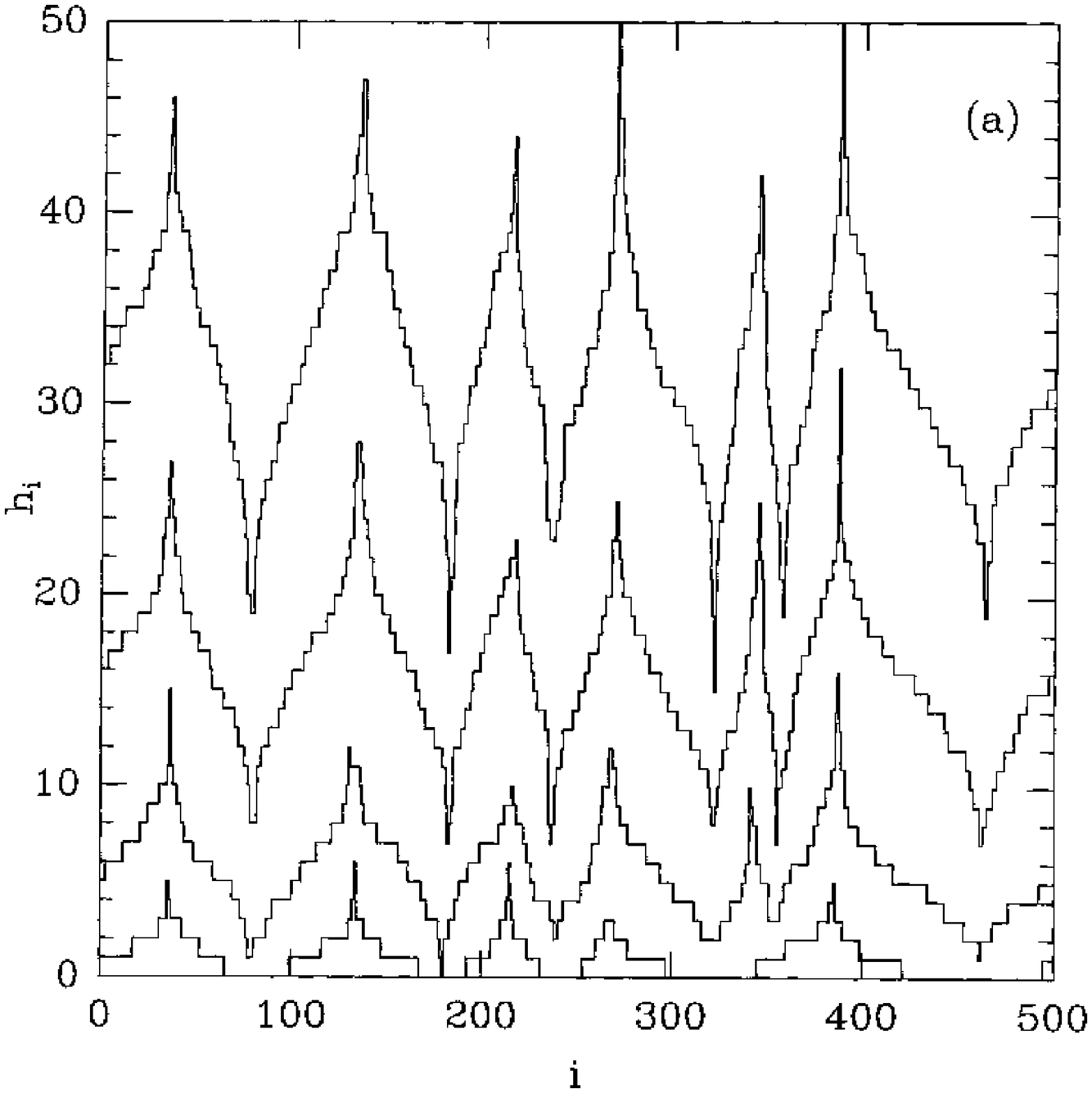} \includegraphics[width=0.495\textwidth]{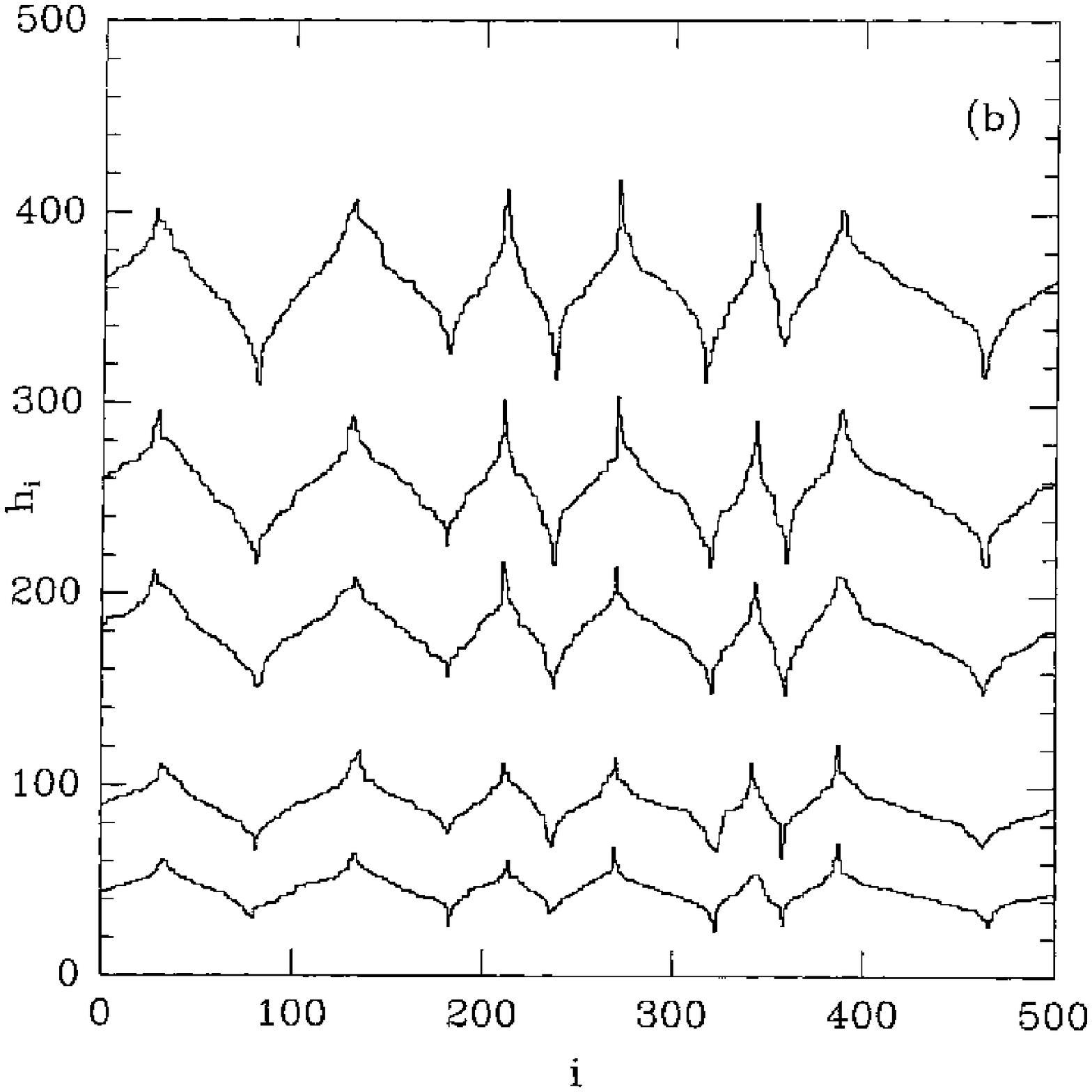} 
\end{center}
\caption[]{Surface morphology in a one-dimensional growth model without
interlayer transport. (a) Morphology after deposition of 1, 5.6, 16
and 32 monolayers. (b) Morphology after deposition
of 45.25, 90.5, 181, 256 and 362 ML.  
The ratio of diffusion to deposition rate
is $D/F = 5 \times 10^6$ \cite{Krug97b}}
\label{wedfig}
\end{figure}

While in-layer diffusion does not affect the 
{\em vertical} surface morphology, as encoded in the layer
coverages $\theta_n$, it is certainly reflected in the
{\em lateral} mass distribution along the surface. This is
illustrated in Figure \ref{wedfig} by a one-dimensional simulation.
It shows the emergence of a fairly regular pattern of 
mound-like surface features with a characteristic pointed shape. 
Each mound consists of a tapering stack of islands upon islands,
reminiscent of a {\em wedding cake}. In the following
we argue that many properties of this pattern
follow immediately from the expression (\ref{Poisson}) for the
layer coverages \cite{Krug97b}.

Figure \ref{wedfig} clearly demonstrates that the lateral positions and sizes
of the mounds originate in the growth of the first layer: The
wedding cakes grow on the templates of the first layer islands, and
their spacing is simply determined by the density $N$ of first layer
nuclei, which in one dimension is proportional to $(F/D)^{1/4}$ 
(see Sect.\ref{NucRW}). The persistence of the first layer pattern
throughout the deposition of hundreds and thousands of layers requires,
first, that no additional mounds nucleate during the later stages
of growth, and, second, that neighboring mounds do not merge. The
first requirement is met because the step spacing on the sides of the
mounds decreases with increasing coverage (the mounds steepen), and
therefore nucleation on the vicinal terraces
making up the sides becomes highly unlikely. The merging of mounds is suppressed
because the lateral positions of the maxima and minima of the surface
profile are strongly correlated from one layer to the next. For the maxima
this reflects the fact that nucleation occurs only in the top layer 
of each mound, while for the minima it can be attributed to the
{\em Zeno effect}, the appearance of deep crevices between 
neighboring mounds \cite{Elkinani94}. 
Such a crevice closes very slowly, because fewer and fewer deposited
atoms find their way to its bottom terrace. In particular,
according to (\ref{Poisson}) the exposed fraction
of the substrate $\varphi_0 = 1 - \theta_1 = e^{-\Theta}$ approaches zero only
asymptotically, but does not vanish at any finite time. 

The resulting picture of the pattern forming process is readily 
generalized to growth on real, 
two-dimensional substrates.
The nucleation of the first layer islands partitions the substrate
into capture zones. Each zone supports a single
mound, which is fed by the atoms deposited into the zone. This implies
that the typical shape
of the mounds can be read off from the layer
distribution (\ref{Poisson}): The area $A_n$ of the $n$'th
layer of a mound will be equal to $\theta_n A_0$, where $A_0$ is
the area of the corresponding capture zone. A more transparent form
of (\ref{Poisson}) is obtained in the limit of thick layers, $\Theta \gg 1$,
when the Poisson distribution (\ref{Poisson2}) can be replaced by a 
Gaussian of width $\sqrt{\Theta}$.
The layer distribution then follows by integration, 
\begin{equation}
\label{Travel}
\theta_n = \Phi((n-\Theta)/\sqrt{\Theta})
\end{equation}
where 
\begin{equation}
\label{Error}
\Phi(s) =  
1 - \frac{1}{\sqrt{2 \pi}} \int_{-\infty}^s dy \; e^{-y^2/2} =
\frac{1}{2}[1 - {\rm erf}(s/\sqrt{2})]
\end{equation}
and ${\rm erf}(s)$ denotes the error function.
Equation (\ref{Travel}) shows that the mounds attain a time-independent
limiting shape when rescaled vertically by $W = \sqrt{\Theta}$.

\subsection{An improved model}
\label{Asympt}

\begin{figure}
\centerline{\epsfig{figure=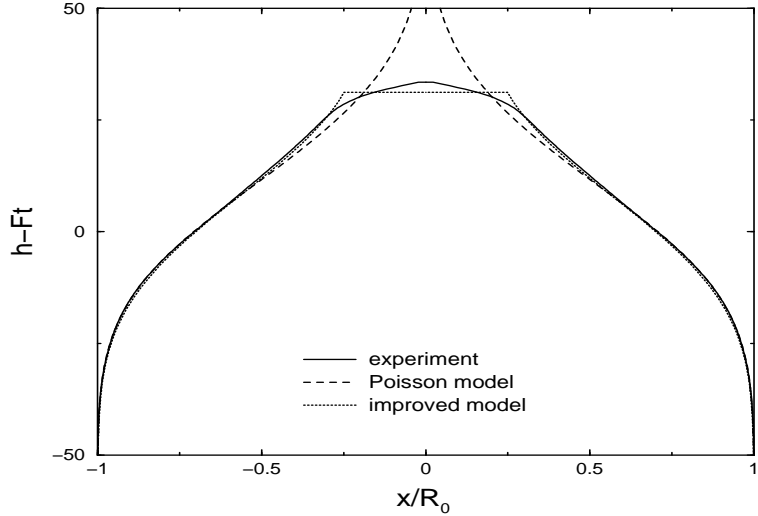,height=10.cm,width=7.cm,angle=-90}}
\caption{Experimentally determined mean mound shape (full line) compared
to the predictions of the Poisson growth model (dashed line), and a fit
to the shape function in the improved
model which includes nucleation on the top
terrace (dotted line). The figure shows a mound composed
of circular islands seen from the side. 
The experimental data are taken from \cite{Kalff99b},
and are courtesy of T. Michely. The experimental conditions
correspond to those of Figure \ref{PtMounds} after deposition of 130 
monolayers.}
\label{moundfit}
\end{figure}

The main features of the Poisson growth model agree well with deposition 
experiments on Pt(111) carried out for film thicknesses ranging
from 0.3 to 300 ML \cite{Kalff99b} (Figure \ref{PtMounds}). 
The mound shape at 130 ML, obtained by averaging the layer coverages
over several mounds, matches the predicted shape
function (\ref{Error}) both in the
valleys and on the slopes of the mounds, but differs significantly
near the tops: Instead of the pointed peaks seen in Figure \ref{wedfig},
the real mounds terminate in flat terraces of a characteristic
lateral size (Figure \ref{moundfit}). 
This discrepancy should be no surprise, since the
model assumption of zero interlayer transport becomes extremely
unrealistic for an adatom that has been deposited onto the freshly
nucleated, small top island of a mound. In the absence of ascending
step edges, such an atom has no choice but to
interrogate the island edge many times
and eventually cross it, even if the crossing probability in each
attempt is very small. As the island grows, the residence time of 
the adatom increases, until the probability for two atoms to be
present simultaneously on the island becomes appreciable, and 
a new nucleus is formed on top of the island.

We now describe a simple modification of the Poisson model which takes
into account the delayed nucleation of the top layer \cite{Krug01}.
A mound is approximated as a stack of concentric,
circular islands. The base is an island of
radius $R_0$ which does not grow, and the radius of the $n$'th island
is $R_n = R_0 \sqrt{\theta_n}$.  
Since there are no sinks for atoms on the top layer, all atoms deposited
there have to attach to the descending step. The top terrace therefore
absorbs all atoms landing on layers $n_{\mathrm{top}}$ and 
$n_{\mathrm{top}} -1$, and grows according to 
\begin{equation}
\label{thetatop}
\frac{d \theta_{n_\mathrm{top}}}{dt} = F \theta_{n_{\mathrm{top}}-1}, 
\end{equation}
while (\ref{Poissongrowth}) still applies for $n < n_{\mathrm{top}}$.
Equations (\ref{Poissongrowth},\ref{thetatop}) have to be supplemented by a rule
for the nucleation of a new top terrace. A simple choice would be to posit
that nucleation occurs whenever the current top layer reaches some critical
coverage. 
More realistically, nucleation should be treated as
a stochastic process governed by the nucleation rate $\omega$ computed in 
Section \ref{2nd}. 
A deterministic rule which is close in spirit
to stochastic nucleation can 
be obtained as follows. Suppose the
current top terrace has nucleated at time $t=0$, and denote its radius
by $R_\mathrm{top}(t)$. Then the probability that no nucleation has
occurred on the top terrace up to time $t$ is given by \cite{Krug00a}
\begin{equation}
\label{P0}
P_0(t) = \exp \left(- \int_0^t dt' \; \omega(R_\mathrm{top}(t'))
\right),
\end{equation} 
and the probability density of the time $t$ of the next nucleation event
is $-dP_0/dt$. It follows that the mean value of $P_0$ at the time of 
nucleation satisfies 
\begin{equation}
\label{P0bar}
\bar P_0 = \int_0^\infty dt \; P_0(t) \left(- \frac{dP_0}{dt} \right) = 
1 - \bar P_0
\end{equation}
and thus $\bar P_0 = 1/2$. In the numerical implementation, we therefore
monitor the increase of 
$P_0$ during the growth of the top terrace and create a new
top terrace when $P_0 = 1/2$. For the nucleation rate 
$\omega$ the expression (\ref{omega}) will be used.

\begin{figure}
\centerline{\epsfig{figure=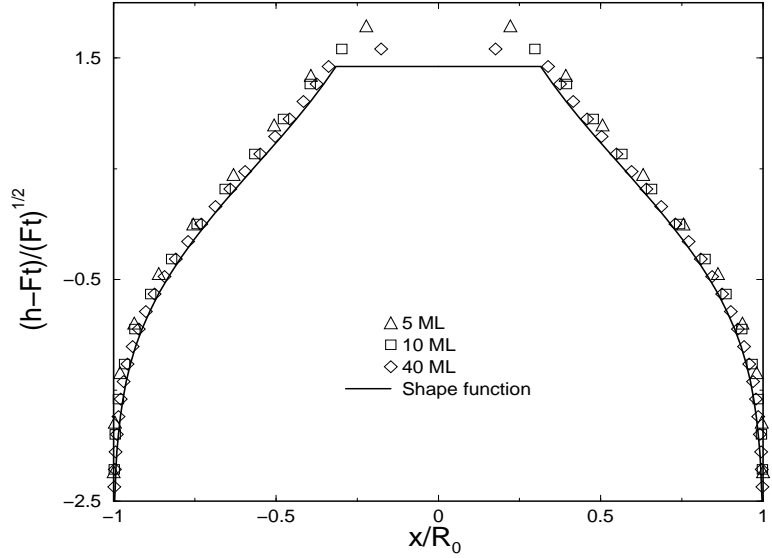,height=10.cm,width=7.5cm,angle=-90}}
\caption{Convergence of numerically generated wedding cakes to the
asymptotic shape.}
\label{cakescale}
\end{figure}

A full analytic solution of the model is difficult
because the nucleation probability (\ref{P0}) depends on the
size of the top terrace, which is determined by the entire
history of the wedding cake through the coupled equations
(\ref{Poissongrowth}). 
The numerical solution of the model equations shows that, like
in the case of Poisson growth,
the height profile of the mound converges to a time-independent
asymptotic shape, when viewed relative to the mean film height
$Ft$ and rescaled horizontally by $\sqrt{Ft}$ (Figure \ref{cakescale}). 
For large coverages this implies a separation of time scales
between the mound sides, which evolve slowly 
at typical step velocities of order $1/\sqrt{Ft}$, and the top terrace, which 
reaches a coverage of order unity during the growth of one layer.
It is therefore reasonable, as a first approximation, to assume that
the former top terrace ceases entirely to grow once
a new island has nucleated on top of it. This should produce
a lower bound on the island radii reached at a given time.

Denote by $R_{n-1}^\mathrm{top}$ the radius of island $n-1$ at
nucleation of island $n$, and by $t_{n}$ the time of
this nucleation event. Then during its tenure as the top
terrace, the radius of island $n$ grows, according to 
(\ref{thetatop}), as 
\begin{equation}
\label{Rntop1}
R_n(t) = \sqrt{F(t-t_{n})} R_{n-1}^\mathrm{top}.
\end{equation}
The time $t_{n+1}$ of the next nucleation event is 
determined by evaluating (\ref{P0}) using (\ref{omega})
and (\ref{Rntop1}), and setting the result
equal to 1/2, 
\begin{equation}
P_0(t_{n+1}) = 
\exp \left( - 
\frac{\pi^2 F^2 (R_{n-1}^\mathrm{top})^5}{2 D'}
\int_{t_n}^{t_{n+1}} dt \; [F(t - t_n)]^{5/2}
\right) = 1/2.
\end{equation}
This implies a recursion relation
\begin{equation}
\label{Rntop}
R_{n}^\mathrm{top} = R_c^{5/7} (R_{n-1}^\mathrm{top})^{2/7}
\end{equation}
for the size of the top terrace at nucleation of the next layer,
where we have introduced the characteristic radius
\begin{equation}
\label{Rc}
R_c = \left( \frac{7 \ln 2}{\pi^2} \right)^{1/5} 
\left( \frac{D'}{F} \right)^{1/5}
\approx 0.868 \cdot \left( \frac{D'}{F} \right)^{1/5}.
\end{equation}
The time interval $\tau_n = t_{n+1} - t_n$ during which
$n_{\mathrm{top}} = n$ is related to  $R_{n}^{\mathrm{top}}$
through (\ref{Rntop1}), and correspondingly satisfies
\begin{equation}
\label{taun}
\tau_{n+1} = F^{-1} (F \tau_n)^{4/7}.
\end{equation}

It is easy to check that the recursion relations (\ref{Rntop})
and (\ref{taun}) approach fixed point values
\begin{equation}
\label{FP}
R_{n}^\mathrm{top} \to R_c, \;\;\;\;
F \tau_n \to 1
\end{equation}
exponentially fast in $n$. Thus asymptotically there is one nucleation
event during the growth of one monolayer, and nucleation occurs when
the radius of the top terrace has reached the value $R_c$.
The numerical solution shows  
that these statements remain valid for the full
dynamics, although the approach of $R_{n}^\mathrm{top}$ and
$\tau_n$ to their asymptotic values is slower than exponential due
to the coupling to the lower layers (the deviations decay as
$1/\sqrt{Ft}$).

To derive the asymptotic mound shape 
analytically, we insert the ansatz (\ref{Travel})
into (\ref{Poissongrowth}) and expand for large
$\Theta$. We find that the shape function $\Phi(s)$ has to satisfy
the differential equation
\begin{equation}
\label{Phi}
\Phi''(s) = - s \Phi'(s).
\end{equation} 
This shows that the inflection point of the profile, where
$\Phi'' = 0$, is always located at $s = 0$, i.e. at 
$n = Ft$. The solution of (\ref{Phi}) which satisfies
the boundary condition $\lim_{s \to -\infty} \Phi(s) = 1$
reads
\begin{equation}
\label{error}
\Phi(s) =  
1 - C [1 + \mathrm{erf}(s/\sqrt{2})],
\end{equation}
where $C$
is a constant of integration. The profile is cut off at the
rescaled height $s_\mathrm{max}$ of the top terrace, where
the coverage takes the value $\theta_c = (R_c/R_0)^2$, 
\begin{equation}
\label{xmax}
\Phi(s_\mathrm{max}) = \theta_c.
\end{equation}
Accordingly, the height of the top
terrace above the mean film thickness $Ft$ is 
$s_\mathrm{max} \sqrt{Ft}$. The Poisson shape (\ref{Error}) is 
recovered in the limit $\theta_c \to 0$, $s_{\mathrm{max}} \to 
\infty$, $C \to 1/2$.

The two parameters $C$ and $s_{\mathrm{max}}$ of the shape function
are related by (\ref{xmax}), but to fix both of them, a further
relation is required. This is provided by the normalization 
condition (\ref{Total})
which, using (\ref{Travel}), translates into
\begin{equation}
\label{phinorm}
\int_0^{s_\mathrm{max}} ds \; \Phi(s) = \int_{-\infty}^0 
ds \; (1 - \Phi(s)).
\end{equation}
Together equations (\ref{error},\ref{xmax},\ref{phinorm})
define a family of shape functions parametrized by
$\theta_c$. 
Other features of the mound shape can be easily extracted.
For example,
the surface width (\ref{W}) turns out to be given by
\begin{equation}
\label{W2}
W^2 \approx (1 - \theta_c) Ft
\end{equation}
asymptotically. By fitting the experimental mound shapes to the predicted
shape function, the top terrace size $R_c$ and hence, through (\ref{Rc}),
the interlayer diffusion rate $D'$ can be determined \cite{Krug00a}.
Assuming equal preexponential factors for in-layer and interlayer diffusion,
the fit shown in Figure \ref{moundfit} yields an additional step edge barrier
of $\Delta E_{\mathrm{S}} \approx 0.21 \; \mathrm{eV}$.

\subsection{The growth-induced current}
\label{Current}

The approach of the preceding subsections has provided us with a fairly
accurate description of the shape of individual mounds. However, 
it does not capture global features of the morphology,
such as the spatial organization of the mounds and their size distribution.
Most importantly, in 
many (though not all \cite{Kalff99b}) experiments the mounds are
observed to \emph{coarsen}, i.e. their typical lateral extent increases
with film thickness \cite{Thuermer95,Stroscio95,Tsui96,Zuo97}.
This requires mass transport between mounds, and hence the basic assumption
of the wedding cake model -- that the entire mass deposited within the capture
zone of a mound contributes to its growth -- 
cannot be upheld\footnote{See however Sect.\ref{Noise} for a mechanism
of mound coarsening without lateral mass transport.}. So far the
only theoretical approach for treating coarsening is based on 
phenomenological continuum equations, which will be discussed in detail in the
next section. Here our purpose is to lend some credibility to this approach
by showing how the asymptotic shape of the wedding cakes can be derived 
within the phenomenological framework.        

\begin{figure}
\centerline{\epsfig{figure=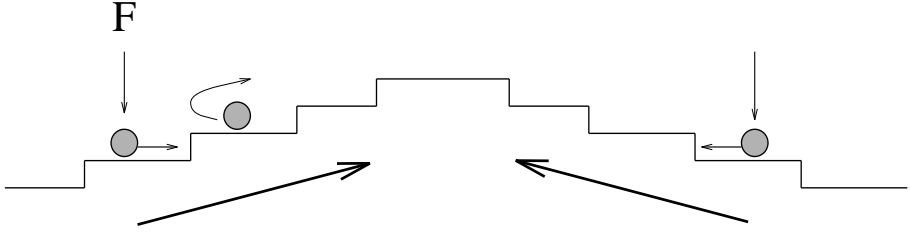,height=3.cm,width=12.cm,angle=0}}
\caption{Illustration of the uphill surface current generated
by step edge barriers.}
\label{currfig}
\end{figure}

The continuum theory of mound formation is based on the notion 
of a growth-induced, uphill surface current 
\cite{Politi00,Krug97a,Villain91,Krug93}. Figure \ref{currfig}
illustrates the basic idea: Because atoms deposited onto 
a vicinal terrace\footnote{A vicinal terrace is a terrace which is 
bounded by an ascending step on one side and a descending step on the 
other. For further discussion of vicinal surfaces see 
Section \ref{Vicinals}.} on the side of a mound attach preferentially
to the ascending step, they travel on average in the direction of the
uphill slope
between the point of deposition and the point of incorporation.
It is important to realize that this does \emph{not} require atoms
to actually climb uphill across steps \cite{Krug97a,Krug93}. 

The evaluation of the current is particularly simple when 
interlayer transport is completely suppressed, 
and nucleation on the vicinal terrace can be 
neglected. An atom deposited
onto a vicinal terrace of width $l$ then travels an average
distance $l/2$ to the ascending step, and hence the current
is $F l/2$. Describing the surface profile by a continuous height 
function  $h({\vec r},t)$ which measures the film thickness
(in units of the monolayer thickness) above a substrate point 
${\vec r}$, the local terrace width is $l = \vert \nabla h 
\vert^{-1}$, and hence the current is given by the expression
\begin{equation}
\label{J}
\vec J = \frac{F}{ 2 \vert \nabla h \vert^2} \nabla h.
\end{equation}
The surface profile evolves according to the continuity
equation
\begin{equation}
\label{cont}
\frac{\partial h}{\partial t} + \nabla \cdot \vec J = F.
\end{equation}
Here we are specifically interested in radially symmetric mounds.
Rewriting (\ref{cont}) in polar coordinates and using (\ref{J})
yields the following evolution equation for the mound profile
$h(r,t)$:
\begin{equation}
\label{hr}
\frac{\partial h}{\partial t} = - \frac{F}{2r}  
\frac{\partial}{\partial r} \; r \left(\frac{\partial h}{\partial r}
\right)^{-1} + F.
\end{equation}
We want to show that (\ref{hr}) possesses solutions corresponding
to the asymptotic mound shape derived in Section \ref{Asympt}.
To this end we make the ansatz
\begin{equation}
\label{hscale}
h(r,t) = \sqrt{Ft} \; \psi(r) + Ft.
\end{equation}
Inserting this into (\ref{hr}) yields the differential equation
\begin{equation}
\label{psi}
\frac{d}{dr} \left( \frac{r}{\psi'(r)} \right)
= r \psi(r).
\end{equation}
In order to establish the equivalence between the shapes
described by $\psi(r)$ and the scaled coverage distribution
$\Phi(s)$ of Section \ref{Asympt}, we need to verify that
the function $\psi(r)$ defined implicitly by
\begin{equation}
\label{equiv}
\left( \frac{r}{R_0} \right)^2 = \Phi(\psi(r)) 
\end{equation}
solves (\ref{psi}). Indeed, taking the derivative with 
respect to $\psi$ on both sides yields
\begin{equation}
\label{equiv2}
r \left( \frac{d\psi}{dr} \right)^{-1} = 
- \frac{C R_0^2}{2} e^{-\psi^2/2}
\end{equation} 
which reduces to (\ref{psi}) upon taking another derivative
with respect to $r$.

This calculation shows explicitly that the sides of the mounds
evolve according to the continuum equation (\ref{cont}); 
in \cite{Krug97b} the same result was obtained for a one-dimensional
geometry. The continuum description does however not include
the nucleation on the top terraces, which has to be added to 
the profile determined by (\ref{psi}) as a boundary condition.
The development of a continuum theory of epitaxial growth which
explicitly incorporates nucleation remains a challenge for the
future.

\section{Phenomenological continuum theory of mound formation}

\label{Continuum}

\subsection{Motivation of the evolution equation}
\label{Motivation}

As it stands,
the continuum equation (\ref{cont}) cannot be used globally,
because the surface current (\ref{J}) diverges for 
$\vert \nabla h \vert \to 0$. For small slopes the nucleation
of islands on the vicinal terraces can no longer be neglected.
In quantitative terms, island nucleation sets in when the width
$l$ of the vicinal terrace becomes comparable to the distance $l_D$ 
between islands nucleated on a flat surface. Using (\ref{Nscal}),
the island distance is estimated as
\begin{equation}
\label{lD}
l_D \approx N^{-1/2} \sim (D/F)^{\chi/2},
\end{equation}
with $\chi = 1/3$ for $i^\ast = 1$. The nucleated islands capture
part of the deposited atoms, thus reducing the flux to the ascending
step, and hence the uphill current. It is easy to show that the resulting net
current vanishes linearly for $\vert \nabla h \vert \to 0$ 
\cite{Politi00,Krug97a}. Ignoring effects of crystal anisotropy (see
Section \ref{Anisotropy}), the current always points in the direction
of $\nabla h$. It is then generally of the form
\begin{equation}
\label{Jgen}
{\vec J}(\nabla h) = f(\vert \nabla h \vert^2) \nabla h,
\end{equation}
where the function $f(u)$ approaches a constant for $u \to 0$.
To match the form (\ref{J}) for large slope, we further require
that $f \sim u^{-1}$ for $u \to \infty$. The transition between the 
two regimes should occur at a slope of the order of $1/l_D$. Under suitable
rescaling \cite{Rost96} this slope, as well as all other dimensionful 
parameters can be set to unity, and the overall behavior of the current
can be represented by the interpolation formula \cite{Johnson94}
\begin{equation}
\label{f1}
f_{\mathrm{I}}(u) = \frac{1}{1 + u}  \;\;\;\;\;(\mathrm{model} \; 
\mathrm{I}).
\end{equation} 

For large slopes the current functions (\ref{J}) and (\ref{f1}) 
decrease indefinitely, but the uphill current remains nonzero for all slopes.
This implies that mass is continuosly transported uphill, leading 
to unbounded steepening of the morphology. Experimentally, it is often
observed that the mound slopes approach a constant, ``magic'' value 
after a transient phase of steepening, a phenomenon known as
\emph{slope selection}. This can be incorporated into
the continuum description by letting the current function vanish
at some nonzero slope \cite{Siegert94}. Scaling the selected
slope to unity, a simple choice of the function $f$ in (\ref{Jgen})
that incorporates slope selection is 
\begin{equation}
\label{f2}
f_{\mathrm{II}}(u) = 1 - u \;\;\;\;\;(\mathrm{model} \; 
\mathrm{II}).
\end{equation} 
The growth equations with current functions (\ref{f1}) and (\ref{f2}) will be 
referred to as model I and model II in the following.

The model defined by Eqs.(\ref{cont},\ref{Jgen}) still does not constitute
a useful description of the growing surface. To understand what difficulty
remains, let us linearize (\ref{cont}) around the flat solution, writing
\begin{equation}
\label{hpert}
h(\vec r,t) = Ft + \epsilon(\vec r,t). 
\end{equation}
We find a diffusion equation
for $\epsilon$, but with a \emph{negative} diffusion coefficient
$- f(0)$ \cite{Villain91}. This confirms that the flat surface is unstable,
but the instability  becomes arbitrarily violent on arbitrarily short length
scales. To cure this unphysical feature, we have to introduce a smoothening
term which counteracts the uphill current on short length scales. 
Near thermal equilibrium, such smoothening is provided by the 
well known Gibbs-Thomson effect, which induces mass transport from 
positively curved regions of the surface (hilltops) to negatively 
curved regions (valleys). To leading order in a gradient expansion,
this implies a mass currrent
\begin{equation}
\label{Jsmooth}
\vec J_{\mathrm{smooth}} = K \nabla (\nabla^2 h),
\end{equation}
where $K$ is proportional to the product of the surface free energy and
the adatom mobility \cite{Mullins59,Krug95}. The Gibbs-Thomson effect
may not be directly relevant under the far-from-equilibrium conditions
of MBE, but other mechanisms have been suggested
that give rise to a smoothening current of the same general form
\cite{Politi00,Tang98,Michely01}. 

We follow the common practice
and add the divergence of Eq.(\ref{Jsmooth}) to the right hand side
of (\ref{cont}).
Repeating the linear stability analysis, we find exponentially growing
or decaying perturbations $\epsilon(\vec r,t) \sim \exp[i (\vec q \cdot
\vec r) + \sigma(\vec q)t]$, where the growth rate $\sigma$ of a 
perturbation of wavevector $\vec q$ is given by
\begin{equation}
\label{sigma}
\sigma(\vec q) = f(0) \vert \vec q \vert^2 - K \vert \vec q \vert^4.
\end{equation}
The instability is now limited to an \emph{unstable band}
$0 < \vert \vec q \vert < q_c \equiv \sqrt{f(0)/K}$. Wavelengths below
$2 \pi /q_c$ are stabilized by the smoothening current (\ref{Jsmooth}).
The most rapidly growing mode, which dominates the initial pattern that emerges
from a generic random perturbation, has wavelength\footnote{If one demands
that the initial wavelength should equal the island spacing
$l_D$, it can be shown that $K \approx F l_D^4$ 
\cite{Krug99}.} $2 \pi \sqrt{2}/q_c$
and a finite growth rate $\sigma(q_c/\sqrt{2})$. The unphysical features
of the instability have thus been successfully removed.

As before, the coefficient $K$ in (\ref{Jsmooth}) can be set to unity
by suitable rescaling. In addition, the constant deposition rate $F$
on the right hand side of (\ref{cont}) can be removed by letting
$h \to h - Ft$. The final outcome of these considerations is the 
evolution equation
\begin{equation}
\label{cont2d}
\frac{\partial h}{\partial t}  = 
- \nabla \cdot [f(\vert \nabla h \vert^2) \nabla h] - 
(\nabla^2)^2 h
\end{equation}
which will be analyzed in the remainder of this section.

\subsection{Driving force of coarsening} 

To gain some insight into the origin of mound coarsening, 
we consider the evolution of the local surface
slope $\vec m = \nabla h$.
The current (\ref{Jgen}) can be written as the negative 
gradient (with respect to ${\vec m}$) of a 
\emph{slope potential} defined by
\begin{equation}
\label{slopepot}
V({\vec m}) = - \frac{1}{2} \int_0^{\vert {\vec m}
\vert^2} du \; f(u).
\end{equation}
The slope potential allows one to introduce a \emph{Lyapunov functional}
\cite{Manneville90}
for the evolution equation (\ref{cont2d}), that is, a functional
of the surface morphology which is a 
strictly decreasing function of time.
Defining
\begin{equation}
\label{Lyap}
\Psi[h({\vec r},t)] = \int d^2{\vec r} \left( \frac{1}{2} (\nabla^2 h)^2 + 
V(\nabla h) \right),
\end{equation}
it is a simple matter to verify that \cite{Siegert97,Moldovan00}
\begin{equation}
\label{Lyap2}
\frac{d \Psi}{dt} = - \int d^2{\vec r} 
\left( \frac{\partial h}{\partial t} \right)^2 < 0.
\end{equation}
This is an extremely useful relation, because it suggests that
the far-from-equilibrium growth process can be viewed in analogy to a
process of relaxation to equilibrium, in which a quantity akin to a
free energy is minimized. 

The two terms in the integrand
of (\ref{Lyap}) describe different aspects
of this minimization: On the one hand, the 
value of the slope potential $V$ should be minimized
locally; on the other hand, the square of the surface curvature 
$\nabla ^2 h$ should become small. The minimization of the 
slope potential drives the process of slope selection. 
It is evident from (\ref{slopepot}) that the
potential has a minimum only if the function $f$ goes through
zero at some nonzero slope; otherwise, $V$ decreases
indefinitely with increasing $\vert {\vec m} \vert$, and the 
attempt to minimize it leads to unbounded steepening of the morphology. 
The minimization
of the surface curvature term in 
(\ref{Lyap}) provides, within the continuum equation 
(\ref{cont2d}), the driving force for coarsening. As will be shown
in the next section, this implies that the coarsening behavior
depends on the spatial distribution of the curvature on the surface.

\begin{figure}
\centerline{\includegraphics[width=0.5\textwidth]{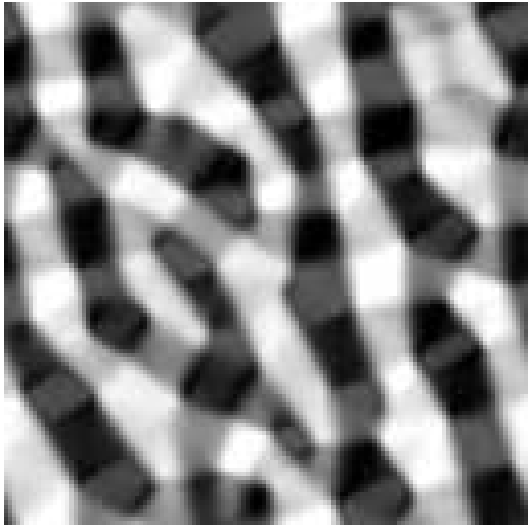}}
\caption{Surface morphology generated by numerical solution of
the growth equation (\ref{cont2d}) with slope selection (model II). 
Within the domains the absolute value of the 
slope is $\vert \nabla h \vert = 1$, and the greyscale encodes
the in-plane direction of the height gradient. Maxima and minima
cannot be told apart, i.e. the morphology is up-down symmetric.
Courtesy of Martin Rost.}
\label{rostfig}
\end{figure}

The analogy of the functional (\ref{Lyap}) 
to a thermodynamic free energy becomes more pronounced
by writing the evolution equation for $\vec m$ in the form
\begin{equation}
\label{modelB}
\frac{\partial {\vec m}}{\partial t} = \nabla \nabla \cdot 
\left( \frac{\delta \Psi}{\delta {\vec m}}\right).
\end{equation}
This is highly reminiscent of the Cahn-Hilliard equation, also
known as model B, describing the phase ordering of a system with
a two-component order parameter $\vec m$ \cite{Bray94}. 
In the case of model II, the order parameter is subject to the
familiar ``Mexican hat'' potential which favors $\vert \vec m \vert = 1$.
Mathematically, (\ref{modelB}) differs from model B in the ordering
of the differential operators in front of the functional derivative
on the right hand side, and in the fact that
${\vec m}$ is irrotational by construction. Physically, 
an important difference is that
the domain boundaries between regions with different orientations 
of the order parameter (i.e., the slope)
are \emph{straight}\footnote{More precisely,
deformations of the domain boundaries are restored on a time scale which
is much shorter than the time scale for coarsening \cite{Siegert97b}.}
(Figure \ref{rostfig}).
This is in contrast to conventional phase ordering kinetics, where
the domain wall curvature drives the coarsening process 
\cite{Siegert97,Bray94}.

\subsection{Coarsening laws} 
To extract the coarsening 
behavior from a nonlinear
continuum equation such as (\ref{cont2d}), one commonly 
imposes a \emph{scaling hypothesis} stating that the patterns at 
different times are similar, in a statistical sense, up to a rescaling
by the average feature size \cite{Bray94}. In the present context
this implies that the height-height correlation function 
(as well as any other statistical measure of
the morphology) depends on time only through the typical lateral mound size
$\lambda$ and the surface width $W$, and hence can be written in the form
\begin{equation}
\label{Gscal}
\langle h(\vec r,t) h(0,t) \rangle = W^2(t) g({\vec r}/\lambda(t)),
\end{equation}
where $g$ is a time-independent scaling function.
In addition, the time dependence of $\lambda$ and $W$ is usually assumed
to follow the power laws
\begin{equation}
\label{power}
\lambda(t) \sim t^n, \;\;\;\;\; W \sim t^\beta,
\end{equation}
which define the \emph{coarsening exponent} $n$ and the
\emph{roughening exponent} $\beta$. The mound slope is then of the order
of $W/\lambda \sim t^{\beta - n}$. Slope selection implies $\beta = n$,
while in the case of steepening $\beta > n$.

To put the scaling hypothesis to work,
we consider the time evolution of the surface width. Since the
mean height has already been subtracted, it is given by 
$W^2 = \langle h^2 \rangle$, where the angular brackets 
represent a spatial average. Multiplying (\ref{cont2d}) by 
$h$ and integrating spatially
we obtain \cite{Rost97b}
\begin{equation}
\label{Wevol}
\frac{1}{2} \frac{d W^2}{dt} = 
\langle f(\vert \nabla h \vert^2) \vert  \nabla h \vert^2
\rangle - \langle (\nabla^2 h)^2 \rangle.
\end{equation}
This clearly demonstrates how the surface is roughened
(the width increased) by the uphill component of the
growth-induced current,
and smoothened by the curvature term in (\ref{cont2d}).
Both terms on the right hand side of (\ref{Wevol}) have
definite signs.
Since the coarsening process can be viewed as a competition between
the two terms, we expect them both, as well as their difference,
to be of a similar order of magnitude\footnote{Exact calculations for 
one-dimensional growth equations show that this assumption may fail,
because the two terms on the right hand side of (\ref{Wevol})
cancel almost completely \cite{Politi98,Politi00c}. 
The simple scaling arguments
presented here then only provide upper bounds on $n$ and 
$\beta$. This kind of behavior seems however to be specific to 
the one-dimensional geometry.}. This assumption is sufficient
to fix the exponents $n$ and $\beta$ entering the scaling
laws (\ref{power}).

According to the scaling hypothesis (\ref{Gscal}),
the typical curvature of the surface is of the order of 
$W/\lambda^2$, and hence the right hand side of (\ref{Wevol})
can be estimated as $W^2/\lambda^4$. This immediately leads to
$n=1/4$ independent of $\beta$.
However, the estimate implicitly assumes that the curvature
is \emph{uniformly} distributed over the surface, which is true
for model I (Eq.(\ref{f1})) without slope selection, but not for 
model II (Eq.(\ref{f2})). In the presence of slope selection, 
numerical integration of the evolution equation (\ref{cont2d}) shows
that the surface breaks up into flat facets at the selected
slope, $\vert {\vec m} \vert = 1$, which are separated by straight
domain boundaries \cite{Moldovan00} (Figure \ref{rostfig}). 
The width of these boundaries introduces another
length scale into the problem, which invalidates the naive application
of the scaling hypothesis \cite{Bray94}. This length scale is independent
of time, and coincides in order of magnitude with the initial 
mound size $2 \pi \sqrt{2}/q_c$. 
The surface curvature vanishes
on the facets and is concentrated in the network of domain boundaries.
The correct estimate of the right hand side of (\ref{Wevol})
is therefore $\langle (\nabla^2 h)^2 \rangle \sim 1/\lambda$, which
yields the scaling relation $2 \beta - 1 = -n$. Since slope
selection also implies $\beta = n$, the result $n=1/3$ follows
for model II. 

To complete the derivation for model I, we note
that because of the unbounded steepening $f_{\mathrm{I}} \approx 
\vert \nabla h \vert^{-2}$ for long
times, and hence the first term on the right hand side of (\ref{Wevol})
is of order unity. Thus $W^2$ increases linearly in time, and $\beta = 1/2$.
In summary, we have
\begin{eqnarray}
\label{modelI}
n = 1/4, \; \beta = 1/2 \;\;\;\;\; (\mathrm{model} \; 
\mathrm{I})
 \\
n = \beta = 1/3 \;\;\;\;\; (\mathrm{model} \; 
\mathrm{II}).
\label{modelII}
\end{eqnarray}
The scaling
arguments for model I can be extended to equations with a smoothening
term of the general form $-(-1)^k (\nabla^2)^k h$ 
with $k > 2$  \cite{Golubovic97}. 
Such terms have been proposed to model
situations in which thermal detachment from steps is impossible,
although no clear identification of the 
associated microscopic processes has
been provided \cite{Stroscio95}. 
The second term on the right hand side of (\ref{Wevol}) then becomes
$\langle (\nabla^k h)^2 \rangle$, which is estimated as 
$W^2/\lambda^{2k}$, leading to $n = 1/2k$. The estimate of the 
first term remains the same as above, so that $\beta = 1/2$
independent of $k$.

\subsection{Crystal anisotropy}
\label{Anisotropy}

Since
the zeros of ${\vec J}$ determine the stable slopes
of the fully-developed mounds, the expression for the current should also
incorporate the crystal symmetry of the surface
\cite{Siegert94}.
To give an example, a possible choice for a surface of square symmetry
reads \cite{Siegert98}
\begin{eqnarray}
  \label{JESsquare}
J_{x} &=& m_x ( 1 - m_x^2 - b m_y^2)  
  \nonumber
\\
J_{y} &=&m_y ( 1 - m_y^2 - b m_x^2),    
\end{eqnarray}
which generates pyramidal mounds with selected slopes 
\begin{equation}
\label{mast}
{\vec m}^\ast = \frac{(\pm 1, \pm 1)}{\sqrt{1 + b}}
\end{equation} 
for
$-1 < b < 1$. This reduces to the isotropic model II for $b = 1$.

The microscopic origin of the anisotropy of 
${\vec J}$ is twofold \cite{Politi00b}.
First, because of its dependence on the step structure through, 
for example, the density of kinks, the effective step edge barrier
depends on the orientation of the step in the plane. Second, 
the growth-induced current contains contributions from the 
diffusion of atoms \emph{along} step edges 
\cite{Pierre-Louis99,Murty99}, which depend even
more strongly on the step orientation. An approximate evaluation  
\cite{Politi00b} of these
contributions shows that it is difficult to make contact
with phenomenological expressions like (\ref{JESsquare}),
because the microscopic analysis predicts -- in contrast to 
(\ref{JESsquare}) -- that the current remains anisotropic
even in the limit of small slopes, ${\vec m} \to 0$.
In mathematical terms, this implies a singularity in the 
function ${\vec J}({\vec m})$ at 
${\vec m} = 0$, which may point to a fundamental problem of the 
continuum theory at small slopes.

An important consequence of crystal anisotropy is the possible
breakdown of the fundamental scaling hypothesis
(\ref{Gscal}) \cite{Siegert98}. 
For the model (\ref{JESsquare}) this occurs due
to the existence of metastable checkerboard mound patterns, which
coarsen only through the motion of dislocations (``roof tops'').
The distance between the dislocations defines a second length scale,
which is much larger than the mound size. 
An analysis of   
the dislocation dynamics  suggests that
$n=1/4$, but for quite different reasons than in the case of model I. 
In the presence of hexagonal crystal
anisotropy Eq.(\ref{modelII}) remains valid, 
because the network of domain boundaries remains 
sufficiently disordered for the scaling arguments
to apply \cite{Moldovan00}.

\subsection{Up-down symmetry and desorption}
As written, Eq.(\ref{cont2d}) is symmetric under the transformation
$h \to -h$, and correspondingly the morphologies it generates are
up-down symmetric, that is, mounds and valleys have the same shapes.
This is in strong disagreement with the mound patterns observed in 
experiments and Monte Carlo simulations, which display isolated mounds 
separated by a connected network of crevices. These morphologies
can be modeled by supplementing (\ref{cont2d}) by a symmetry-breaking
term \cite{Stroscio95}. In a gradient expansion, 
the lowest order term which accomplishes this is of the 
form $\nabla^2 (\nabla h)^2$ \cite{Villain91}.
For a one-dimensional evolution equation, it has been shown that
such a term does not qualitatively change the coarsening behavior 
\cite{Politi98}. Since much of the 
analysis in this lecture relied on the introduction of the 
slope potential $V$ defined in (\ref{slopepot}), which
becomes impossible in the presence of a symmetry-breaking term,
it is unclear if the same is true in two dimensions. 
In fact, it seems that the effect of the symmetry-breaking may be quite
substantial: If the ridges are removed from the network of domain boundaries
and only the crevices remain, it is no longer possible to localize most
of the surface curvature in the domain boundaries, and the scaling arguments
developed above have to be modified. 

\begin{figure}
\centerline{\includegraphics[width=1.0\textwidth]{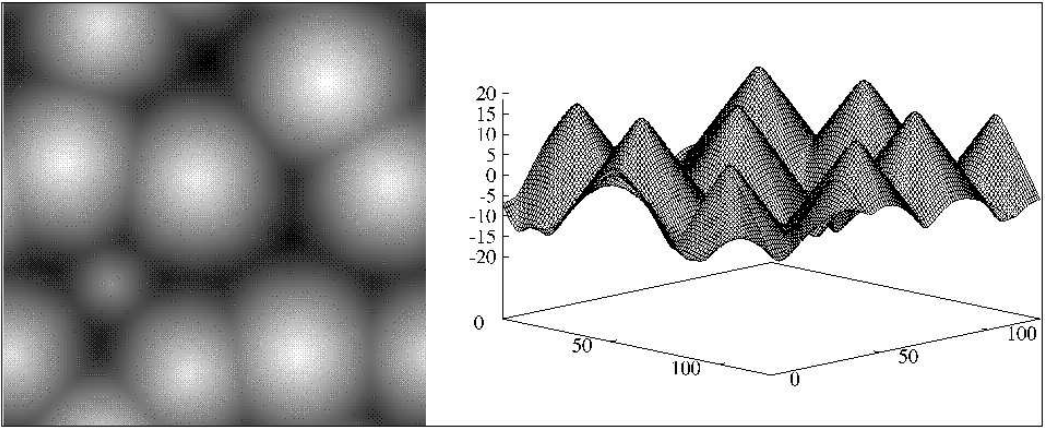}}
\caption{Mound configurations generated by numerical solution of
the evolution equation (\ref{desorb}). Left panel shows a greyscale
representation of the height. Note the distinct
up-down asymmetry \cite{Smilauer99}.}
\label{desorbfig}
\end{figure}

The up-down symmetry is also broken when desorption from the surface is allowed
for. The probability for a deposited atom to redesorb from 
the surface before being
captured at a step evidently depends on the step density, and hence on the
surface slope. The desorption rate is therefore an even function of
$\nabla h$. Adding such a function to the right hand side of (\ref{cont2d})
fundamentally changes the character of the evolution equation, because the
slope-dependent desorption rate cannot be written as the divergence of a 
current. In \cite{Smilauer99} the effect of desorption on mound coarsening
was studied in the framework of the evolution equation 
\begin{equation}
\label{desorb}
\frac{\partial h}{\partial t} = - \nabla \cdot [(1 - (\nabla h)^2)(\nabla h)]
- \frac{\alpha}{1 + (\nabla h)^2}  - (\nabla^2)^2 h,  
\end{equation}
where $\alpha > 0$ is a dimensionless measure of the desorption rate. 
The numerical integration of (\ref{desorb}) shows the emergence of conical
mounds separated by a network of crevices
(Figure \ref{desorbfig}). The form of the desorption term
on the right hand side of (\ref{desorb}) implies that most desorption occurs
from maxima and minima, where $\nabla h \approx 0$. Since the minima form a 
one-dimensional network, while the maxima (the tips of the cones) are 
point-like objects, the growth rate at the minima is smaller than that at the
maxima by an amount of the order of $1/\lambda$, where $\lambda$ is the
lateral mound size. The surface width then increases according to 
$d W/dt \sim 1/\lambda$. Together with the slope selection property of 
(\ref{desorb}) this implies $\beta = n = 1/2$. We conclude
that desorption leads to a 
significant speedup of coarsening.

\subsection{Noise-induced mound coarsening}
\label{Noise}

The theory developed so far has been entirely deterministic.
Here we show that the most important source of fluctuations,
the shot noise in the deposition beam, induces an alternative
coarsening mechanism which generally competes with the 
curvature-driven coarsening described above\footnote{Quantitative
analysis shows, however, that the noise-induced mechanism is negligible
in most published experiments.} \cite{Tang98,Krug97d}. 

\begin{figure}
\centerline{\epsfig{figure=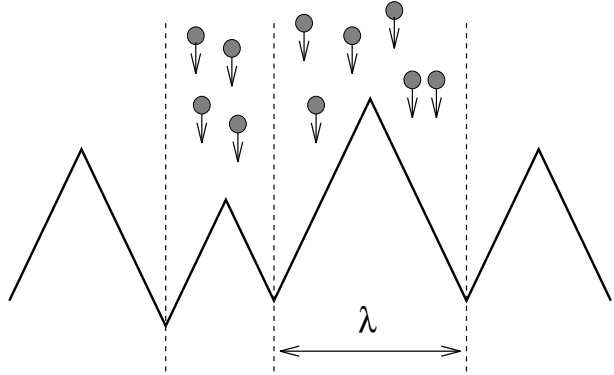,height=5.cm,width=8.cm,angle=0}}
\caption{Illustration of noise-induced mound coarsening.}
\label{noisefig}
\end{figure}

The basic idea is illustrated in
Figure \ref{noisefig}. Consider an array of roughly equal sized mounds
of area $A = \lambda^2$ and height $H = m \lambda/2$, where $m$ is
the mound slope. During a time $t$, a number $F A t$ of atoms is
deposited onto a mound, with a statistical fluctuation of $\pm
\sqrt{F A t}$. If there is no mass transport between neighboring
mounds, this translates into a relative height fluctuation $\delta
H =  \sqrt{FAt}/A$. Coarsening then occurs if, by
chance, a mound overgrows a less fortunate neighbor. The condition
for this to happen is that $\delta H \approx H$, which implies the
coarsening law
\begin{equation}
\label{lambdaNoise} \lambda \approx m^{-1/2} \Theta^{1/4}.
\end{equation}
Under conditions of slope selection, $m = const.$, the coarsening
exponent\footnote{For a $d$-dimensional substrate,
$n = 1/(d + 2)$.} is $n = 1/4$, while in general the exponent
relation
\begin{equation}
\label{exprel} \beta + n = 1/2
\end{equation}
follows.
This expresses a competition between coarsening and roughening (or
steepening): The larger $\beta$, the smaller $n$, with the
limiting case $\beta = 1/2$, $n = 0$ corresponding to the case of
Poisson growth discussed in Section \ref{Wedding}.

\section{Growth on stepped surfaces}

\label{Vicinals}

The orientation of a vicinal surface is close to (in the
\emph{vicinity} of) a high symmetry direction of the crystal
lattice. Such a surface therefore consists of several lattice
spacings wide, high index terraces separated by steps, usually of
monolayer height.
During growth on a vicinal surface, the attachment of freshly
deposited adatoms to the preexisting steps competes with
nucleation of islands on the terraces. Nucleation is expected to
be negligible if the distance $l$ between the preexisting steps is
small compared to the island spacing $l_D$ \cite{Pimpinelli97}
(see also Sect.\ref{Motivation}). The surface then maintains its
vicinal shape, and growth occurs through step propagation or
\emph{step flow}. In the following the conditions for step flow
will be assumed to hold\footnote{Due to the stochastic
nature of nucleation, the question about the ultimate stability
of step flow is somewhat subtle \cite{Politi00}. For the one-dimensional
Poisson growth model of Sect.\ref{Poissonmodel}, it has been shown that
step flow is always \emph{metastable}, and estimates for
the time scale at which it breaks down have been derived 
\cite{Krug95b}. A similar breakdown has been seen in two-dimensional
simulations \cite{Rost96}, but the underlying mechanism is not
clear.}.

On a perfectly ordered vicinal surface, as it would appear in
thermal equilibrium at low temperatures, the steps are straight
and equally spaced. Correspondingly, the morphological
instabilities of stepped surfaces are of two kinds: Either the
individual steps develop a \emph{meander}, beyond their thermal or
kinetic roughness, or several steps form \emph{step bunches},
regions of high step density separated by large terraces.
The main topic of this lecture is a generic step meandering
instability in homoepitaxial growth, which was first predicted
theoretically by Bales and Zangwill \cite{Bales90}. It is caused
by the asymmetry between ascending and descending steps which the
step edge barrier introduces. Meandering instabilities
which have been attributed to the Bales-Zangwill mechanism
were identified experimentally 
on surfaces vicinal to Pt(111) \cite{Rost96} and
Cu(100) \cite{Schwenger97,Maroutian99,Maroutian01}. Step bunching during
homoepitaxy has been observed on several semiconductor surfaces
\cite{Schelling99,Tejedor99}, but a simple generic mechanism has
not been suggested.

\subsection{Stability of a Step Train}
\label{StepTrainStability}

\begin{figure}[htb]
\centerline{\includegraphics[width=0.6\textwidth]{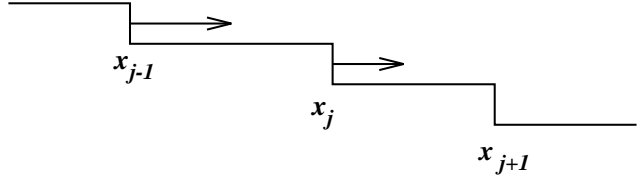}}
\caption[]{Schematic of a growing step train.}
\label{StepTrainFig}
\end{figure}

From a theoretical perspective, the step flow growth mode is attractive
because it can be described in terms of step motion without the
need to treat island nucleation. The problem simplifies further if
the steps are assumed to be straight. Then the propagation speed
of the $j$-th step in a step train can be written as the sum of
the contributions $f_-$ and $f_+$ from the upper and lower
terraces, each of which is a function of the corresponding terrace
width (Fig.\ref{StepTrainFig}). Denoting the position of the
$j$-th step by $x_j$, the evolution equations then read
\begin{equation}
\label{StepTrain} \frac{d x_j}{dt} = f_+(x_{j+1}-x_j) +
f_-(x_j-x_{j-1}).
\end{equation}

A train of equally spaced steps moving at speed $v_0 =
f_+(l)+f_-(l)$ evidently satisfies these equations. To probe its
stability, we consider a small perturbation of the form
\begin{equation}
\label{TrainStab} x_j(t) = j l + v_0 t + \epsilon_j(t)
\end{equation}
and linearize (\ref{StepTrain}) in the $\epsilon_j$. The solutions
of the linearized equations are of the form $\epsilon_j(t) \sim
\exp[i \phi j + \sigma(\phi)t]$, where the growth rate $\sigma$ is
given in terms of the phase shift $\phi$ by the expression
\begin{equation}
\label{omegaphi} \sigma(\phi) = -(1 - \cos \phi)(f_+'(l) -
f_-'(l)) + i \sin \phi (f_+'(l) + f_-'(l)).
\end{equation}
Stability requires the real part of $\sigma$ to be negative for
all $\phi$, which implies
\begin{equation}
\label{stability} \frac{d}{dl} (f_+(l) - f_-(l)) > 0.
\end{equation}
Roughly speaking, (\ref{stability}) expresses the fact that a step
train is stable if the steps are fed primarily from the lower
terrace, in the sense that $f_+ > f_-$ \cite{Schwoebel66}. This is
easy to understand intuitively: Under this condition a step
trailing a particularly wide terrace accelerates, and the uniform
step spacing is restored. When (\ref{stability}) is violated the
step train is unstable towards step bunching. The largest growth
rate is then attained for $\phi = \pi$, hence step pairs form in
the initial stage of the instability\footnote{This need no longer
be true if long ranged step-step interactions are taken into
account.}.

While the above analysis applies generally to growing or
sublimating vicinal surfaces, we now specialize to a surface
growing in the absence of evaporation. The straightforward
evaluation of the functions $f_\pm$ \cite{Elkinani94} 
then shows that a
growing step train is stable whenever $D' < D$. A step
bunching instability during growth would require a ``negative''
step edge barrier, in the sense of $D' > D$. Similar arguments
show that normal step edge barriers generically do cause step
bunching during sublimation \cite{Schwoebel66,Schwoebel69}.

The stabilization of the equidistant step train by the step edge
barrier may be interpreted in terms of an effective,
growth-induced step-step repulsion. This repulsion is very
efficient, in the sense that the resulting terrace width
fluctuations can be far smaller than in thermal equilibrium
\cite{Krug95b,Pierre-Louis98}.

\subsection{The Bales-Zangwill instability}
\label{Meander}

\begin{figure}[htb]
\centerline{\includegraphics[width=0.8\textwidth]{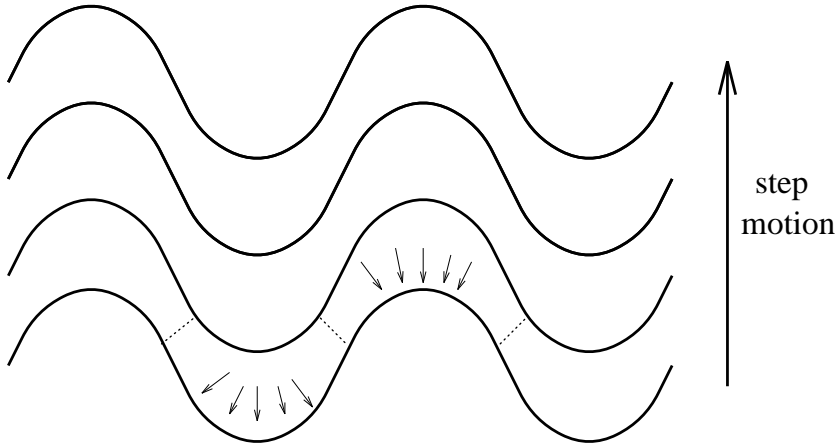}}
\caption[]{Schematic of the Bales-Zangwill mechanism for step
meandering} \label{MeanderFig}
\end{figure}

Bales and Zangwill made the remarkable observation that the very
same mechanism that stabilizes a growing vicinal surface against
step bunching also makes the steps susceptible to a meander
instability \cite{Bales90}. Figure \ref{MeanderFig} illustrates
the phenomenon on a qualitative level. To account for the mutual
repulsion between the steps, it is assumed (and will be confirmed
by the quantitative analysis) that they meander in phase. The
terraces can then be subdivided into lots, as indicated by the
dotted lines. Each lot receives the same number of atoms per unit
time, which attach primarily to the corresponding segment of the
ascending step. Because of the meander, the indented segments of
the step are longer than the protruding ones. Since both capture
the same flux, the protrusions propagate faster and the
deformation is amplified.

\begin{figure}[htb]
\centerline{\includegraphics[width=0.8\textwidth]{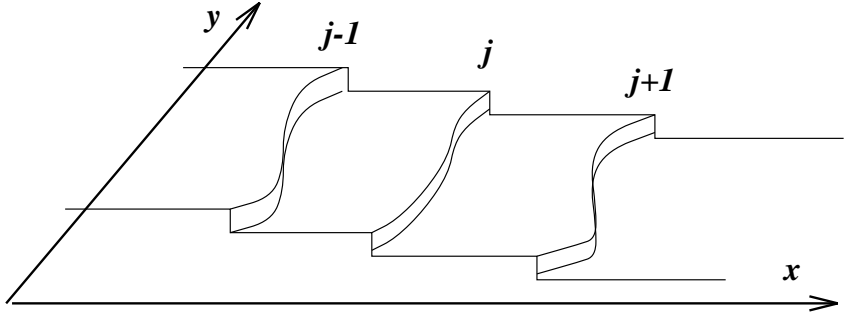}}
\caption[]{Geometry of the vicinal surface used in the stability
analysis.} \label{BCFFig}
\end{figure}

For the quantitative stability analysis, we use the coordinate
system shown in Fig.\ref{BCFFig}. The position of the $j$-th step
is described by a function $\zeta_j(y,t)$. The step spacing
of the unperturbed surface is $l$. Between the steps 
the adatom density $n({\vec r},t)$ satisfies a diffusion equation,
which we employ in the stationary form 
\begin{equation}
\label{Fickstat}
D \nabla^2 n + F = 0.
\end{equation}
This is justified provided the time scale for the motion of a step
across a terrace, $1/F$, is large compared to the diffusion time
scale $l^2/D$, i.e. if the \emph{P\'eclet number} \cite{Ghez88}
\begin{equation}
\label{Pe}
\mathrm{Pe} = F l^2/D 
\end{equation}
is small compared to unity. Since the step spacing 
in step flow growth 
has to be small compared to $l_D$, which in turn is small compared
to $(D/F)^{1/2}$ because of the relation (\ref{lD}), the stationarity
condition is always satisfied\footnote{Step motion beyond the 
stationary approximation is treated in \cite{Ghez88}.}. 

The attachment and detachment of 
adatoms at the steps is described by the boundary conditions
\cite{Schwoebel69,Ghez88}  
\begin{equation}
\label{bc}
\pm D {\vec {\mathbf n}} \cdot \nabla n 
\vert_{x = \zeta_j \pm 0} = 
k_\pm (n - n_{\mathrm{eq}} \vert_{x=\zeta_j \pm 0}).
\end{equation}
Here $k_+$ and $k_-$ denote the attachment rates to an ascending and
a descending step, respectively, $\mathbf n$ is the normal vector of 
the step, and $\vec n_{\mathrm{eq}}$ is the equilibrium adatom density
at the step.
The thermodynamic cost of step deformations enters
the boundary conditions through the expression
\begin{equation}
\label{neqdeform} n_{\mathrm{eq}} = n_{\mathrm{eq}}^{(0)} \left( 1
+ \frac{\tilde \gamma \kappa_{\mathrm{st}}}{k_{\mathrm{B}} T}
\right),
\end{equation}
where $\tilde \gamma = \gamma + d^2 \gamma/d \vartheta^2$ is the
step stiffness, related to the orientation-dependent
step free energy 
$\gamma(\vartheta)$, 
$n_{\mathrm{eq}}^{(0)}$ is the equilibrium adatom density at a
straight step, and $\kappa_{\mathrm{st}}$ 
is the step curvature. 
Equation (\ref{neqdeform}) is the two-dimensional analog
of the Gibbs-Thomson relation mentioned in Sect.\ref{Motivation}.

Once the boundary value problem defined by
(\ref{Fickstat},\ref{bc},\ref{neqdeform}) has been
solved for a given configuration of steps, the local normal
velocity $v_n^{(j)}$ of each step can be computed from the total
mass flux reaching the step from the two adjoining terraces, as
well as through diffusion \emph{along} the step. This yields
\begin{equation}
v_n^{(j)} = D \left[{\vec {\mathbf n}} \cdot \nabla n
\vert_+ - {\vec {\mathbf n}} \cdot \nabla n \vert_- \right] +
\frac{\partial}{\partial s} \mu_{\mathrm{st}} 
\frac{\partial}{\partial s} \tilde \gamma
\kappa_{\mathrm{st}}^{(j)},
\end{equation}
where $\mu_{\mathrm{st}}$ is the mobility
for migration along a curved (that is, atomically rough) step \cite{Krug95},
and $s$ denotes the arc length of the step. 

The calculation now proceeds, in principle, as in 
Sect.\ref{StepTrainStability}. The general form of the perturbed
step train is
\begin{equation}
\label{StepTrain2d}
\zeta_j(y,t) = j l + v_0 t + \epsilon_j(y,t).
\end{equation}
To linear order in the $\epsilon_j$, the solution of the coupled
equations can be decomposed into normal modes of the form
$\epsilon_j(y,t) \sim \exp[i \phi j + i q y + \sigma(\phi,q) t]$.
Here $q$ denotes the wavenumber of the step deformation, corresponding
to a meander wavelength $2 \pi/q$. The real part of
the growth rate $\sigma(\phi,q)$
turns out to be maximal for the in-phase mode
$\phi = 0$ \cite{Pimpinelli94}. This is a consequence of the kinetically
induced step repulsion described in Sect.\ref{StepTrainStability}: The
in-phase mode is a compromise which allows the deformed steps to keep
the terrace width as uniform as possible. 

Since the incipient morphology
will be dominated by the fastest growing mode, we may assume $\phi = 0$
in the following. Moreover, for the present purposes it is sufficient
to consider long wavelength deformations, with a meander wavelength
large compared to the mean step spacing $l$. In this limit
the expression for the growth rate then reads \cite{Gillet00}
\begin{equation}
\label{BZdisp}
\sigma(0,q) = \frac{F l^2 f_s}{2} q^2 - \left(
\frac{D n_{\mathrm{eq}}^{(0)} l}{k_{\mathrm{B}} T} + 
\mu_{\mathrm{st}}\right) \tilde \gamma q^4.
\end{equation} 
The positive
term proportional to $q^2$ describes the destabilization of the 
straight step by the attachment asymmetry. The strength of the destabilization
is proportional to the flux $F$, and to the 
factor 
\begin{equation}
\label{fs}
f_s = \left ( \frac{k_{+}-k_{-}}{k_+ k_- \ell/D +k_{-}+k_{+}}\right)   
\end{equation}
which is a dimensionless measure of the strength of the step edge barrier.
The negative term proportional to $q^4$ describes the
thermal relaxation of the step towards the (straight) equilibrium shape.
The smoothening is driven by the step stiffness $\tilde \gamma$, and it
operates through two kinetic channels \cite{Pimpinelli93b}: 
Detachment-reattachment processes over the terrace, with a rate proportional
to the terrace diffusion coefficient and the terrace width, and 
step edge diffusion with a rate proportional to $\mu_{\mathrm{st}}$. 
    
The form of the growth rate (\ref{BZdisp}) is the same that was derived
in Sect.\ref{Motivation} for the early stages of the mound
instability (Eq.(\ref{sigma})), and correspondingly the physics is very similar.
For sufficiently small $q$ the quadratic term in (\ref{BZdisp}) wins over
the quartic term, and hence the step is subject to a long wavelength
instability for arbitrarily small flux\footnote{In the presence of desorption,
which is the case originally considered by Bales and Zangwill \cite{Bales90},
the instability sets in only above a critical flux.}. The range of unstable
wavelengths is bounded below by $2\pi/q_c$, where $q_c$ is 
the wavenumber at which the two terms on the right
hand side of (\ref{BZdisp}) balance. 
The dominant meander wavelength $\lambda_{\mathrm{BZ}}$ 
corresponds to the maximum
of (\ref{BZdisp}), which yields $\lambda_{\mathrm{BZ}} = 2 \pi  \sqrt{2}/q_c$.
Explicitly, 
\begin{equation}
\label{BZlambda}
  \lambda_{\mathrm{BZ}} = 4 \pi \sqrt{\frac{
(D n_{\mathrm{eq}}^{(0)} l/k_{\mathrm{B}} T + 
\mu_{\mathrm{st}}) \tilde \gamma }
{F l^2 f_s}}.
\end{equation}

\subsection{Nonlinear step meandering}

An analytic approach to the evolution of the meander instability beyond the
linear regime has been developed by Misbah, Pierre-Louis and coworkers
\cite{Gillet00,Pierre-Louis98b}. Since the
in-phase mode is the most unstable
according to linear stability analysis,
the two-dimensional surface morphology can
be represented by a one-dimensional function $\zeta(y,t)$ describing
the displacement of the common step profile from the flat straight
reference configuration $\zeta = 0$. A solvability condition arising
from a multiscale expansion in $\mathrm{Pe}^{1/2}$ then yields 
the evolution equation 
\begin{equation}
\label{nonlinear}
\zeta_t = -\left\{ \frac{\alpha \zeta_y}{1+\zeta_y^2}  +
\left(\frac{\beta}{1+\zeta_y^2} + 
\frac{\beta'}{\sqrt{1 + \zeta_y^2}} \right)
\left[ \frac{\zeta_{yy}}{(1+\zeta_y^2)^{3/2}}
\right]_y \right\}_y. 
\end{equation}
Here subscripts denote derivatives, and the values of the coefficients
$\alpha$, $\beta$ and $\beta'$ can be read off by comparison with
(\ref{BZdisp}). The expression in square brackets
is the step curvature, and the terms proportional to 
$\beta$ and $\beta'$ 
describe step smoothening through attachment/detachment kinetics
and step edge diffusion, respectively. The form of these terms
follows from simple geometric considerations \cite{Kallunki00}.
The two terms differ by 
a factor of $\sqrt{1 + \zeta_y^2}$, because the attachment/detachment
kinetics depends on the step width, while step edge diffusion does
not (compare to (\ref{BZdisp})).

Two types of analytic solutions to (\ref{nonlinear}) have been 
found \cite{Pierre-Louis98b,Kallunki00}.  
{\it Stationary} solutions are obtained by setting the mass 
current along the step (the quantity inside to curly brackets 
on the right hand side of (\ref{nonlinear})) to zero. In terms
of $m(y) = \zeta_y/\sqrt{1 + \zeta_y^2}$ the
stationarity condition reduces to the 
oscillatory motion of a classical particle in a potential,
which can be solved by quadratures.
One thus obtains a one-parameter family of periodic profiles 
$\zeta_S(y)$ which
are most conveniently parametrized by the maximum slope
$S \equiv \max_y \zeta_y$.
We discuss separately the special cases $\beta' = 0$
and $\beta = 0$. For $\beta' = 0$, the amplitude $A(S)$ 
of the profile is an increasing
function of $S$, while the wavelength $\Lambda(S)$ decreases
with increasing $S$, starting out at $\Lambda(0) = \lambda_c 
= 2 \pi/q_c$.
For $S \to \infty$ finite limiting values $A(\infty) = 
\sqrt{8 \beta/\alpha}$, $\Lambda(\infty) =
\sqrt{2 \pi \beta/\alpha} \; \Gamma(3/4)/\Gamma(5/4) \approx 
0.5393527..\lambda_c$ are approached. In contrast, for $\beta = 0$
the potential is harmonic, hence the wavelength of the stationary
profile is $\lambda_c$ independent of its amplitude.   

The \emph{separable} solutions of 
interest read
\cite{Krug97b,Pierre-Louis98b}
\begin{equation}
\label{wed}
\zeta(y,t) = 2 \sqrt{\alpha t}\  \mathrm{erf}^{-1} \left( 
1 - 4 \vert y \vert/\lambda \right), \;\;\;
-\lambda/2 < y < \lambda/2,
\end{equation}
where the wavelength $\lambda$ is arbitrary. 
Equation (\ref{wed}) solves (\ref{nonlinear}) exactly in the limit
$t \to \infty$, when the smoothening terms on the right hand side
becomes negligible compared to the first term, and the evolution
equation reduces to $\zeta_t = -(\alpha/\zeta_y)_y$,
the one-dimensional version of the wedding cake equation
discussed in Sect.\ref{Current}.
The solution (\ref{wed}) is highly singular near the maxima
and minima, where it diverges as $\zeta \sim \pm \sqrt{\ln(1/
\vert y - y_0 \vert)}$, $y_0 = 0, \pm \lambda/2$. 

\begin{figure}[htb]
\centerline{\includegraphics[width=0.7\textwidth]{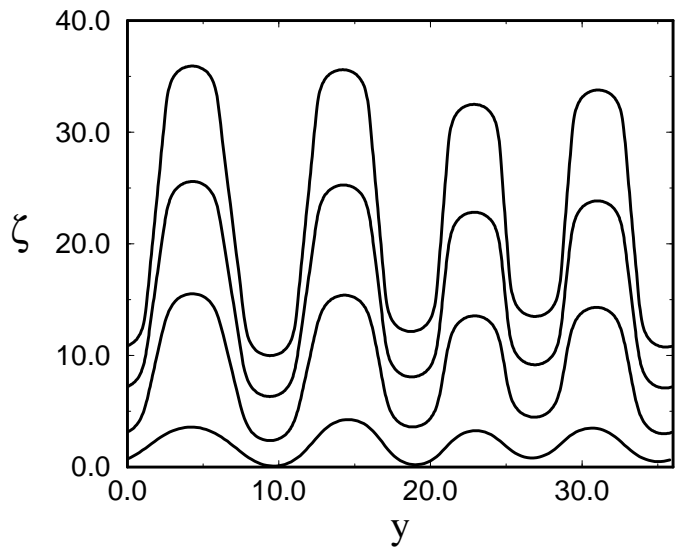}}
\caption[]{Evolution of the step profile starting from a flat
initial condition with small random
fluctuations, for the case of pure step edge diffusion
($\beta = 0$). Subsequent profiles have been shifted
in the $\zeta$-direction. The
$y$-axis has been scaled by $\lambda_c$.} 
\label{rand_prof}
\end{figure}

In Figure \ref{rand_prof} we show results of a numerical integration
of (\ref{nonlinear}), starting 
from a small amplitude random initial condition.
A regular meander pattern
of wavelength $\lambda_{\mathrm{BZ}}$ develops, with an amplitude growing
indefinitely as $\sqrt{t}$. Closer inspection reveals that the 
sides of the profile follow the separable solution; however, the singular
spikes at the maxima and minima of 
(\ref{wed}) are replaced by smooth {\it caps} 
which consist of pieces of the stationary solutions. Since the slope of 
(\ref{wed}) increases monotonically upon approaching an extremum while
it decreases for the stationary profiles, the matching of the two 
solutions occurs near the point of maximum slope. For $t \to \infty$
the slope of the separable solution diverges, and therefore the cap
profile approaches the limiting stationary solution $\zeta_\infty(y)$,
and the length of the cap becomes $\Lambda(\infty)$. The rescaled step
profile $\zeta(y,t)/\sqrt{t}$ approaches an invariant shape in which
the cap appears as a flat facet.

\subsection{Competing instability mechanisms}

The first quantitative experimental test of the predictions of 
Bales and Zangwill was carried out by Ernst and collaborators
for surfaces vicinal to Cu(100) \cite{Maroutian99,Maroutian01}.
Based on the experimentally determined dependence of the meander
wavelength on temperature and flux, they concluded that their
results were \emph{not} consistent with the prediction 
(\ref{BZlambda}). In particular, the observed flux dependence
$\lambda_{\mathrm{meander}} \sim F^{-0.2}$ differs considerably
from the predicted $\lambda_{\mathrm{BZ}} \sim 
F^{-1/2}$ \cite{Maroutian01}. 

This finding suggests that a mechanism different from the 
one described by Bales and Zangwill may be responsible for
the step meandering on Cu(100). A plausible alternative
was proposed in \cite{Pierre-Louis99}, where it was pointed out
that a one-dimensional analog of the mounding instability
discussed in Sections \ref{Wedding} and \ref{Continuum} should
occur on a step, if the diffusion of step adatoms across 
``descending'' kinks
were suppressed by additional energy barriers.
If such barriers are sufficiently strong, a one-dimensional
wedding cake morphology should develop along the step, with a 
characteristic length scale given by the distance $l_{\mathrm 1d}$
between the one-dimensional nuclei forming on the straight
step in the initial stages of growth.  This length scale can be 
estimated from one-dimensional nucleation theory (see Sect.\ref{NucRW}).
In order of magnitude, $l_{\mathrm{1d}} \sim 
(D_{\mathrm e}/F_{\mathrm{1d}})^{1/4}$, where $D_{\mathrm{e}}$ is the
coefficient of one-dimensional diffusion along a \emph{straight}
(rather than kinked) step edge, and $F_{\mathrm{1d}} = F l$ is the 
effective flux impinging onto a unit length of the step edge. A more
precise calculation yields \cite{Politi97}
\begin{equation}
\label{l1d}
l_{\mathrm{1d}} \approx \left( \frac{12 D_{\mathrm{e}}}{F l } 
\right)^{1/4}.
\end{equation}
In contrast to (\ref{BZlambda}), this expression is consistent with
the experimental observations. 

\begin{figure}[htb]
\centerline{\includegraphics[width=0.8\textwidth]{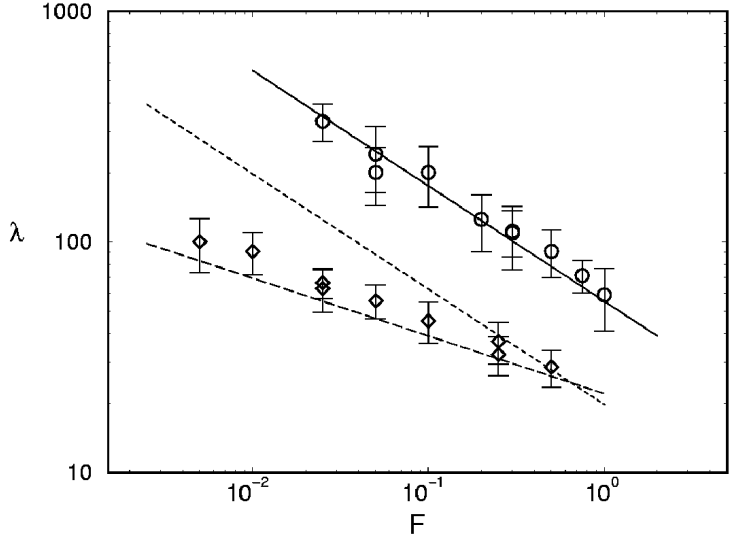}}
\caption[]{Meander wavelength as a function of flux for a simple
cubic solid-on-solid growth model. Diamonds and circles refer to conditions
with and without facilitated edge diffusion, respectively. The full and
dashed lines are the predictions of (\ref{BZlambda}) and (\ref{l1d}) 
for the two sets of parameters, while the dotted line shows the Bales-Zangwill
prediction for the case of facilitated edge diffusion. For details on 
how the material parameters entering (\ref{BZlambda}) and (\ref{l1d})
are evaluated for the solid-on-solid model, see \cite{Kallunki01}.} 
\label{lengths}
\end{figure}

Detailed kinetic Monte Carlo simulations
of stepped Cu(100) surfaces support the idea of a destabilization of 
the steps by kink barriers, and show wedding cake-like step profiles
\cite{Rusanen01}. Simulations of a simple cubic SOS-model have moreover
demonstrated that both instability mechanisms, the Bales-Zangwill
scenario and the kink barrier scenario, can be observed in the
same system by tuning the energy barrier for the detachment of atoms
from steps \cite{Kallunki01}. When detachment is suppressed compared to
edge diffusion, the kink barrier mechanism prevails, while the Bales-Zangwill
instability is realized when detachment rates are comparable to edge
diffusion rates. In both cases quantitative agreement with the predictions
(\ref{BZlambda}) and (\ref{l1d}) can be achieved without adjustable
parameters (Figure \ref{lengths}).

\section*{Acknowledgements}

I would like to thank my collaborators Jouni Kallunki, Mirek Kotrla,
Philipp Kuhn, Thomas Michely, Paolo Politi, Martin Rost and 
Pavel \v{S}milauer for 
fruitful and enjoyable interactions. This work has been supported by
DFG within SFB 237 \emph{Unordnung und grosse Fluktuationen}, and 
by Volkswagenstiftung.

\end{document}